\documentclass[
	aip,
	amsmath,amssymb,
	reprint
]{revtex4-2}

\usepackage[pdftex]{
	graphicx,
	xcolor,
	hyperref
}

\usepackage{dcolumn} % Align table columns on decimal point
\usepackage{bm} % bold math

\usepackage{CJK} % CJK is a legacy package and incompatible with LuaLaTeX
\usepackage[utf8]{inputenc}
\usepackage[T1]{fontenc}

\usepackage{etoolbox} % for \patchcmd command
\usepackage{textgreek} % This package enables not only \textmu command but the correct (i.e., non-italic) print of \um command of the siunitx package.
\usepackage{natmove} % citation should be after punctuation

\usepackage{siunitx}
\newcommand{\nnnegf}{nextnano.NEGF}
\newcommand{\schroedinger}{Schr\"odinger }
\newcommand{\kp}{$\mathbf{k}\cdot\mathbf{p}$ }
\newcommand{\imag}{\mathrm{i}}
\newcommand{\spaced}[1]{\mathop{}\!\mathrm{d}{#1}\mathop{}}
\newcommand{\ee}[1]{\mathop{}\,\mathrm{e}^{#1}}
\newcommand{\kBTe}{k_\mathrm{B}T_\mathrm{e}}
\newcommand{\quotes}[1]{``#1''}
\newcommand{\polarizationVector}{\hat{\mathbf{e}}}

\newif\ifFlatFolderStructure
\FlatFolderStructuretrue

\makeatletter
\def\@email#1#2{%
 \endgroup
 \patchcmd{\titleblock@produce}
  {\frontmatter@RRAPformat}
  {\frontmatter@RRAPformat{\produce@RRAP{*#1\href{mailto:#2}{#2}}}\frontmatter@RRAPformat}
  {}{}
}%
\makeatother

\begin{document}
\title[
	Impact of carrier injector design on the threshold of interband cascade lasers
]{
	Impact of carrier injector design on the threshold of interband cascade lasers
}

\begin{CJK}{UTF8}{ipxm}
\author{Takuma \surname{Sato} (佐藤拓磨)}

	\email{takuma.sato@nextnano.com}
	\affiliation{nextnano GmbH, Konrad-Zuse-Platz 8, 81829 Munich, Germany}
	\affiliation{TUM School of Computation, Information and Technology, Technical University of Munich, Hans-Piloty-Straße 1, 85748 Garching, Germany}
\author{Borislav \surname{Petrovi\'{c}}}
	\affiliation{Julius-Maximilians-Universit\"{a}t W\"{u}rzburg, Physikalisches Institut, Lehrstuhl f\"{u}r Technische Physik, Am Hubland, 97074 W\"{u}rzburg, Germany}
\author{Robert \surname{Weih}}
	\affiliation{nanoplus Advanced Photonics Gerbrunn GmbH, Oberer Kirschberg 4, 97218 Gerbrunn, Germany}
\author{Fabian \surname{Hartmann}}
\author{Sven \surname{H\"{o}fling}}
	\affiliation{Julius-Maximilians-Universit\"{a}t W\"{u}rzburg, Physikalisches Institut, Lehrstuhl f\"{u}r Technische Physik, Am Hubland, 97074 W\"{u}rzburg, Germany}
\author{Stefan \surname{Birner}}
	\affiliation{nextnano GmbH, Konrad-Zuse-Platz 8, 81829 Munich, Germany}
\author{Christian \surname{Jirauschek}}
	\affiliation{TUM School of Computation, Information and Technology, Technical University of Munich, Hans-Piloty-Straße 1, 85748 Garching, Germany}
\author{Thomas \surname{Grange}}
	\affiliation{nextnano Lab, 12 chemin des prunelles, 38700 Corenc, France}

\date{\today}

\begin{abstract}
	% [Objective] 
	We theoretically investigate how the injector region design of interband cascade lasers (ICLs) impacts the threshold carrier and current densities.
	% [Method]
	The model combines a polarization-sensitive 8-band \kp calculation, electrostatics, and a microscopic calculation of Auger recombination rates. The inelastic carrier-carrier scattering is included to lowest order using quasi-equilibrium Green's functions.
	It captures the combined effects of charge-carrier redistribution, parasitic absorption, and bias voltage on the Auger recombination rate.
	% [Result]
	We show that heavily doping the electron injector suppresses the dominant multi-hole Auger recombination by reducing the hole population of the recombination quantum wells. This agrees with the experimental observation that the heavy doping reduces threshold currents.
	Unlike the measurements, however, they do not increase at high doping concentrations in our model, which does not include scattering-mediated carrier escape and/or light absorption.
	Furthermore, by introducing indium to the conventional $\mathrm{Ga}\mathrm{Sb}$ hole injector wells, we explain the rule of thumb from experiments that raising the hole injector levels does not outperform the doping strategy.
	% [Conclusion]
	Our model provides physical insights for optimizing ICL carrier injectors.
\end{abstract}

\maketitle

\end{CJK}

\section{Introduction}
\label{sec:introduction}

Gas sensing applications require portable, room-temperature light sources in the mid-infrared spectral range.
Thanks to the continuous improvement of their design~\cite{Meyer2020} since the first experimental realization by Yang in 1997~\cite{Lin1997}, interband cascade lasers (ICLs) realized to date consume less input power than diode lasers and quantum cascade lasers, operating above room temperature. ICLs are therefore attractive for environmental monitoring~\cite{Wysocki2007}, industrial process control~\cite{Edlinger2015}, clinical breath analysis~\cite{Henderson2018}, and telecommunications~\cite{Delga2019}. 

Theoretical analysis has played a central role in the design of ICL cascade stages.
In the dawn of the development, Meyer et al.\ proposed the \quotes{W-shaped} quantum wells (W-QWs) for radiative recombination in mid-infrared interband lasers, demonstrating with an 8-band \kp calculation that the type-II QWs improve the electrical confinement and suppress Auger recombination compared to type-I band alignments~\cite{Meyer1995}.
The recombination QWs have been subsequently optimized at various emission wavelengths to enhance their oscillator strength \cite{Ryczko2013,Janiak2013,Motyka2015,Motyka2016_GaAsSb,Dyksik2017,Petrovic2025_VQW} and to reduce parasitic valence intersubband absorptions \cite{Hedwig2022,Nauschuetz2023_VBabsorption,Petrovic2025_VQW}. A correlation has been observed between calculated parasitic absorption and measured threshold current densities. 
Recombination QWs may further be optimized in terms of Auger suppression; for InAs-based type-II superlattices, Grein et al.~\cite{Grein1995,Grein2002_article} proposed to engineer the miniband dispersion by strain.

While theoretical predictions about ICLs have mainly focused on the recombination QWs, injector engineering has been experimentally proven to be effective for reducing threshold current densities. For instance, Vurgaftman et al.~\cite{Vurgaftman2011,Bewley2012} achieved \qty{134}{\ampere\per\cm\squared} by heavily $n$-doping the electron injector wells. Experiments~\cite{Vurgaftman2011,WeihPhD} suggest an optimal doping concentration of \qty{5e18}{\per\cm\cubed}. While Vurgaftman et al.\ attributed it to a rebalancing of the electron and hole concentrations in the recombination region, the connection between carrier balance and Auger rates remains unclear due to the limitations of phenomenological models. 
Despite successive experimental refinements in the injector design~\cite{Olafsen1998,Bauer2010}, a comprehensive understanding is still lacking as to how and why the injector design affects the threshold current~\cite{Meyer2020}.
Answering these questions requires a microscopic calculation of Auger recombination.

In this paper, we systematically calculate how the injector design impacts threshold carrier and current densities. The threshold is found through a bias-dependent 8-band \kp gain calculation considering the entire ICL period and parasitic absorptions (subsection \ref{subsec:threshold_carrier_densities}). Without relying on the phenomenological Auger coefficients, we calculate Auger recombination rates from the $GW$ self-energy (subsection \ref{subsec:Auger_rate}). Sec.\,\ref{sec:results_and_discussion} discusses the correlation of the injector designs with the carrier balance and threshold current density. Sec.\,\ref{sec:conclusion} concludes the analysis and outlines future perspectives.
SI units are used throughout this work.

\section{Method}
\label{sec:method}

\subsection{Cascade design}
\label{subsec:method_design}
\begin{figure}
    \centering
	\ifFlatFolderStructure
		\includegraphics[width=\linewidth]{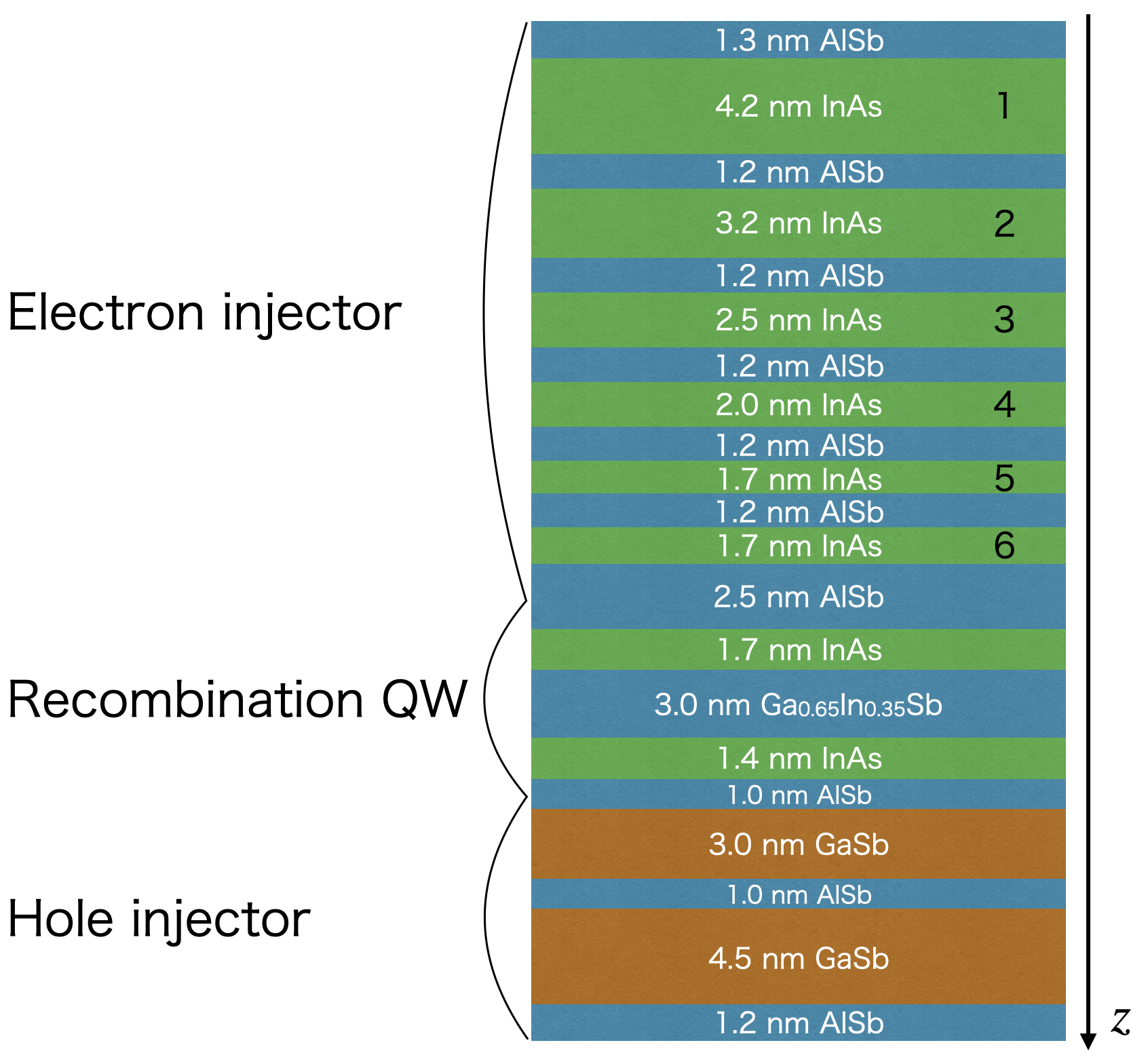}
	\else
		\includegraphics[width=\linewidth]{./images/paper/ICL_period_Vurgaftman.png}
	\fi
    \caption{
		Design of the ICL period in this study. Numbers enumerate the QWs to be doped.
	}
	\label{fig:ICL_layers_in_period}
\end{figure}

We take the ICL structures of Vurgaftman et al.~\cite{Vurgaftman2011} as an example (Fig.\,\ref{fig:ICL_layers_in_period}). The ICLs emit photons of wavelengths from \qtyrange{3.6}{3.9}{\um} and their electron injector consists of 6 quantum wells. We vary the injector designs as listed in Table~\ref{table:design_variations}. The threshold current densities of the reference and some of the 2- and 4-well doping designs were measured experimentally~\cite{Vurgaftman2011,WeihPhD,VurgaftmanPrivateComm}.
The volume doping density is denoted by $y$~(\unit{\per\cm\cubed}).
The doping series can be compared on an equal footing by the sheet doping density in one period $N_\mathrm{D}$~(\unit{\per\cm\squared}).
The $z$ axis is set along the growth direction.

\begin{table}[]  %[tb]
	\caption{
		Doping profile of the electron injector (in $10^{18}$~\unit{\per\cm\cubed}) and QW material for the hole injector among the design variations in this study. $y$ is the doping concentration within each electron injector QW. The value for the indium series has been chosen to maintain the same sheet doping density as the reference.
	}
	\label{table:design_variations} 
	\begin{ruledtabular}
		\begin{tabular}{l|llllll|l}
			& \multicolumn{6}{l}{Electron injector doping} & Hole injector \\
			Design & 1 & 2 & 3 & 4 & 5 & 6 &  \\\hline
			Reference & 0.4 & 0.4 & 0.4 & 0.4 & 0 & 0 & $\mathrm{Ga}\mathrm{Sb}$ \\
			2-well & 0 & 0 & 0 & $y$ & $y$ & 0 & $\mathrm{Ga}\mathrm{Sb}$ \\
			4-well & 0 & $y$ & $y$ & $y$ & $y$ & 0 & $\mathrm{Ga}\mathrm{Sb}$ \\
			Indium & 0 & 0.506 & 0.506 & 0.506 & 0.506 & 0 & $\mathrm{Ga}_{1-x}\mathrm{In}_x\mathrm{Sb}$ \\
		\end{tabular}
	\end{ruledtabular}
\end{table}

\subsection{Threshold carrier densities}
\label{subsec:threshold_carrier_densities}
\begin{figure}
    \centering
	\ifFlatFolderStructure
		\includegraphics[width=\linewidth]{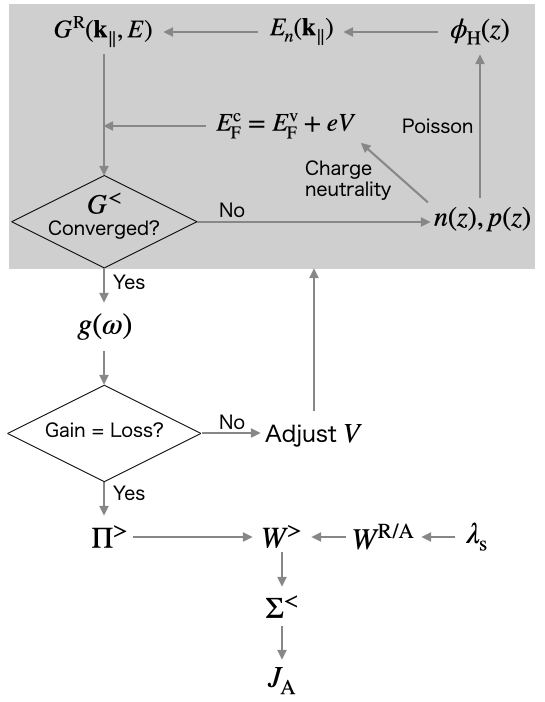}
	\else
		\includegraphics[width=\linewidth]{./images/paper/flowchart.png}
	\fi
    \caption{
		Simulation flow for calculating the threshold carrier densities and Auger recombination current.
	}
	\label{fig:flowchart}
\end{figure}

\begin{figure}
    \centering
	\ifFlatFolderStructure
		\includegraphics[width=\linewidth]{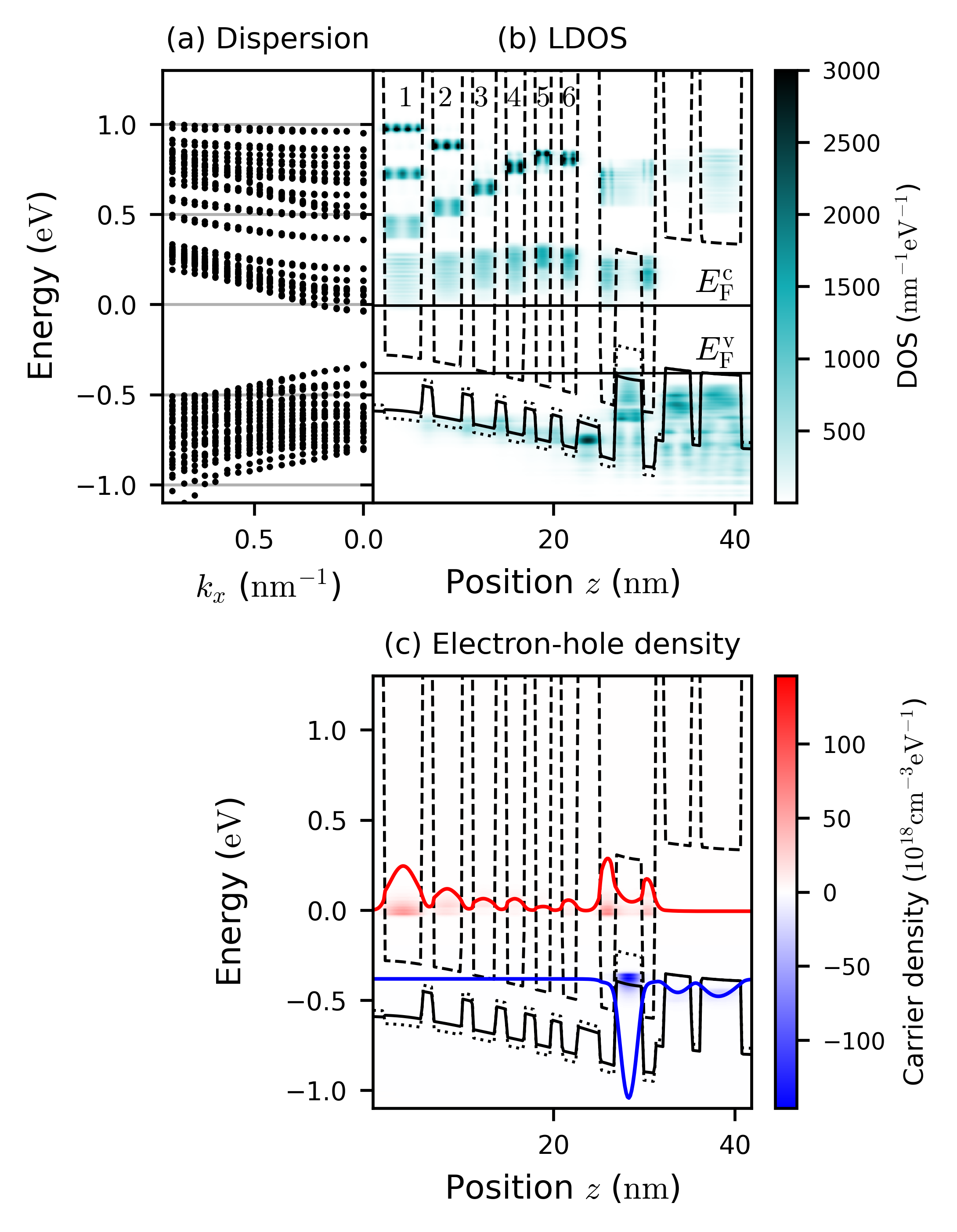}
	\else
		\includegraphics[width=\linewidth]{./images/paper/nextnanopy/doping2D_4.76e11_P6_375.384mV_95TE_GW_NVBO_axial1000500DOS_and_carrier.png}
	\fi
    \caption{
		One cascade stage of the reference ICL in Table~\ref{table:design_variations}
		at the potential drop per period of \qty{375}{\mV}.
		(a) In-plane dispersion in the entire reciprocal space included in the simulations. 
		(b) LDOS integrated over the in-plane momentum. The horizontal lines indicate the quasi-Fermi levels. The dashed, dotted, and solid curves represent the conduction, heavy-hole, and light-hole bandedges of the bulk materials, respectively.
		(c) Energy-resolved electron and hole densities in quasi-equilibrium (colormap). Red and blue curves show the relative magnitude of the energy-integrated electron and hole densities, respectively.
	}
	\label{fig:ICL_states_densities}
\end{figure}

To model the intricacy of threshold carrier densities, i.e., the interplay of carrier distribution, electrostatics, and parasitic absorptions, we conduct the simulations sketched in Fig.\,\ref{fig:flowchart}. The key features are that (i) the carrier distribution in an entire cascade period is solved self-consistently with Poisson's equation, (ii) material gain and parasitic absorption are calculated in the period, (iii) the lasing thresholds are found considering the optical losses outside of the cascade structure.
Since we focus our analysis on the lasing threshold, we may neglect stimulated emission. In this regime, interband transitions in the W-QW are much slower than the relaxation inside carrier injectors, distribution between the injectors and W-QW, and the tunneling through the semimetallic interface. 
Indeed, measured threshold currents~\cite{Vurgaftman2011,WeihPhD,VurgaftmanPrivateComm} and carrier densities calculated below translate to the upper laser state lifetimes ranging from \qtyrange{0.5}{1}{\ns}. They are much longer than the gain recovery time of a few picoseconds measured for \qty{3.3}{\um} ICLs~\cite{Pilat2025}. This justifies assuming constant Fermi levels for the electrons and holes, which are energetically split by the potential drop per period (\textit{quasi-equilibrium approximation}~\cite{ChowKoch_SemiconductorLaser}), see Fig.\,\ref{fig:ICL_states_densities}(b). 
As the carriers relax to the lattice temperature in picoseconds, the carrier temperature $T_\mathrm{e}$ in the Fermi--Dirac functions can be set to the lattice temperature $T_\mathrm{L} = \qty{300}{\kelvin}$.

To calculate the carrier distribution and gain, we have extended the quantum transport simulator \nnnegf~\cite{Grange2014} based on the non-equilibrium Green's functions (NEGF) formalism, to interband devices.
As described below, we adopt here a simplified Green's function method where (i) line broadening is included phenomenologically and (ii) the lesser Green's function, $\bm{G}^<$, describing the energy-resolved electron density, is derived in the quasi-equilibrium approximation. A more comprehensive study including (i) microscopic calculation of intraband scattering processes and (ii) tunneling and leakage currents will be numerically more involved and is left for further work.
The 2-band models, implemented into interband NEGF transport solvers in previous works~\cite{Aeberhard2019,Kolek2023_T2SL}, are not sufficient for the present study; they do not describe the anticrossings of the valence subbands, leading to inaccurate calculation of the parasitic valence intersubband absorptions in ICLs. Thus, we have implemented the 8-band \kp Hamiltonian in Appendix~\ref{app:Hamiltonian}. The material parameters used are summarized in Tables~\ref{table:material_parameters} and \ref{table:material_parameters_bow}. 

At the beginning of a simulation, energy eigenstates $\{|\phi_n\rangle\}$ and subband dispersions of given heterostructure are found without bias voltage at tens of in-plane wavevectors.
The in-plane anisotropy of the band structure~\cite{Pei2004} is not expected to affect the subband occupation significantly in the quasi-equilibrium limit and is therefore neglected.
Second, we select conduction (valence) subbands by the energy range of $E_\mathrm{cutoff}^\mathrm{c}$ ($E_\mathrm{cutoff}^\mathrm{v}$) from their respective ground states and discard other subbands. The in-plane wavevector is considered up to the point where the lowest conduction subband rises by $E_\mathrm{cutoff}^\parallel$. 
The actual cut-off energies used in the simulations are summarized in Appendix~\ref{app:sim_parameters}.
Fig.\,\ref{fig:ICL_states_densities}(a) shows the in-plane dispersion of the selected subbands. Comparison to the bandedge diagram in Fig.\,\ref{fig:ICL_states_densities}(b) confirms that these energies cover the range relevant for the optical transitions and the Auger recombination described in Sec.\,\ref{subsec:Auger_rate}. Third, the selected eigenstates are used to obtain the reduced matrix representation of the position operator $(z_\mathrm{red})_{mn} = \langle \phi_m| \hat{z} |\phi_n\rangle$. The latter is diagonalized to obtain localized modes, which we refer to as the reduced real space (RRS) basis functions~\cite{Grange2014,LeeWacker2006}, and corresponding eigenpositions $\{z_\alpha\}$. The basis gives the matrix representation $\bm{H}_\mathrm{RRS}(\mathbf{k}_\parallel)$ of the Hamiltonian. The RRS modes are enumerated by the indices $\alpha, \beta, \dots$.

Once the RRS basis is constructed, we start the Dyson--Poisson iteration (shaded box in Fig.\,\ref{fig:flowchart}). The electrostatic potential excluding the bias voltage, $\phi_\mathrm{H}(z)$, is calculated from Poisson's equation. Here, the periodic boundary condition is imposed on $\phi_\mathrm{H}(z)$ and $\mathrm{d} \phi_\mathrm{H}(z)/\mathrm{d} z$. The electrostatic potential is incorporated into the Green's functions as follows.
To obtain the retarded Green's function, we solve the eigenvalue problem
\begin{equation}
	\left\{
		\bm{H}_\mathrm{RRS} 
		- e
		\left[
			\frac{V}{l} \bm{z}_\mathrm{red} + \phi_\mathrm{H}(\bm{z}_\mathrm{red})
		\right]
	\right\}
	|\phi_n^\mathrm{red}\rangle
	=
	E_n(\mathbf{k}_\parallel)
	|\phi_n^\mathrm{red}\rangle,
\end{equation}
where $V$ and $l$ are the potential drop per period and period length, respectively.
The retarded Green's function is diagonal in this energy eigenbasis and given by Dyson's equation,
\begin{equation}\label{eq:Dyson_eq_equil}
	G_{nn}^\mathrm{R}(\mathbf{k}_\parallel, E) 
	=
	\frac{\mathcal{N}}{E - E_n(\mathbf{k}_\parallel) - \Sigma_{nn}^\mathrm{R}(\mathbf{k}_\parallel, E)},
\end{equation}
where $\mathcal{N}$ is the normalization factor. We model the line broadening due to elastic scattering by the retarded self-energy,
\begin{equation}
	\Sigma_{nn}^\mathrm{R}(\mathbf{k}_\parallel, E) 
	=
	-\frac{\imag \gamma_\mathrm{subband} / 2}{1 + \left[\frac{E - E_n(\mathbf{k}_\parallel)}{\gamma_\mathrm{subband}}\right]^2},
\end{equation}
where $\gamma_\mathrm{subband}$ is fixed at \qty{15}{\meV}.
Scattering due to interface roughness and charged impurities is not included microscopically in the present study, as their coupling to the $GW$ self-energy would require a self-consistent loop.
LO phonon scattering is not calculated here; the mechanism is fast and accelerates the thermalization of charge carriers, justifying the quasi-equilibrium approximation. 
When Eq.\,\eqref{eq:Dyson_eq_equil} is transformed to the RRS basis, the equal-position spectral functions, $A_{\alpha\alpha}(\mathbf{k}_\parallel, E) = -2\,\mathrm{Im}\,G_{\alpha\alpha}^\mathrm{R}(\mathbf{k}_\parallel, E)$, give the local density of states (LDOS) shown in Fig.\,\ref{fig:ICL_states_densities}(b).
Our Green's functions are in the \quotes{electron picture} and thus do not rely on the concept of holes; selected conduction and valence subbands are first emptied entirely, and the lesser Green's function describes the state occupancy by electrons, regardless of whether the energy is above or below the band gap. In quasi-equilibrium, it is given in the basis $\{|\phi_n^\mathrm{red}\rangle\}$ as
\begin{equation}\label{eq:Glesser_quasi_equilibrium}
	G_{nn}^<(\mathbf{k}_\parallel, E)
	=
	\begin{cases}
		\imag f^\mathrm{c}(E) A_{nn}(\mathbf{k}_\parallel, E) \quad (E > E_\mathrm{b}) \\
		\imag f^\mathrm{v}(E) A_{nn}(\mathbf{k}_\parallel, E) \quad (E < E_\mathrm{b})
	\end{cases}
\end{equation}
where $f^\mathrm{c/v}(E) = \left\{1 + \exp{\left[(E - E_\mathrm{F}^\mathrm{c/v})/(\kBTe) \right]}\right\}^{-1}$ are the Fermi--Dirac distribution functions, $E_\mathrm{F}^\mathrm{c/v}$ are the quasi-Fermi levels for the conduction and valence bands, respectively, and $E_\mathrm{b}$ is the energy border located within the heterostructure band gap.
Although our calculation is performed in the \quotes{electron picture}, we convert the data to the electron-hole densities in Sec.\,\ref{sec:results_and_discussion} as shown in Fig.\,\ref{fig:ICL_states_densities}(c). The population inversion in the W-QW generates the gain.
The iteration in the shaded box in Fig.\,\ref{fig:flowchart} continues until the lesser Green's function and the electrostatic potential converge, and the quasi-Fermi levels satisfy the overall charge neutrality.

After the Dyson--Poisson convergence, we extract level populations and calculate the optical gain spectrum under the dipole approximation using Fermi's golden rule~\cite{Chuang},
\begin{align}
	g(\polarizationVector, \omega)
	&=
	\frac{\pi e^2}{\hbar c \epsilon_0 \sqrt{\epsilon(\omega)}}
	\frac{1}{V}
	\sum_{n>m} 
	\sum_{\mathbf{k}_\parallel}
	\notag\\
	&\quad
	\times
	O_{nm}(\polarizationVector, \mathbf{k}_\parallel)
	\left[
		P_n(\mathbf{k}_\parallel)
		-
		P_m(\mathbf{k}_\parallel)
	\right] \notag\\
	&\quad
	\times
	\mathcal{L}
	\left(
		E_n(\mathbf{k}_\parallel) - E_m(\mathbf{k}_\parallel) - \hbar\omega
	\right), \label{eq:gain_formula_Fermi}
\end{align}
where $e\ (>0)$, $c$, $\epsilon_0$, $\epsilon(\omega)$, and $\mathbf{k}_\parallel = (k_x, k_y)$ are the elementary charge, speed of light, vacuum permittivity, (spatially-averaged) dielectric function, and in-plane wavevector, respectively. The indices $m$ and $n$ enumerate the initial and final subbands. As the summation includes all subbands in Fig.\,\ref{fig:ICL_states_densities}(a), Eq.\,\eqref{eq:gain_formula_Fermi} contains interband, conduction intersubband, and valence intersubband transitions.
The oscillator strengths $O_{nm}$ are calculated from the light polarization vector $\polarizationVector$ and the 8-band momentum operator $\hat{p}_j(\mathbf{k}_\parallel)$ as described in Appendix~\ref{sec:oscillator_strengths}. 
$E_n(\mathbf{k}_\parallel)$ are the eigenvalues of $\bm{H}_\mathrm{RRS}$ plus bias.
The subband populations $P_n$ are calculated from the converged lesser Green's functions transformed to the energy eigenbasis,
\begin{equation}
	P_n(\mathbf{k}_\parallel)
	=
	-\imag
	\int \frac{\spaced{E}}{2\pi}
	G_{nn}^<(\mathbf{k}_\parallel, E).
\end{equation}
The energy broadening is taken to be a Gaussian
\begin{equation}
	\mathcal{L}(E)
	=
	\frac{1}{\sqrt{2\pi}\gamma_\mathrm{gain}}
	\exp{\left( - \frac{E^2}{2\gamma_\mathrm{gain}^2} \right)}.
\end{equation}
For simplicity, we fix the FWHM of all transitions to $2\sqrt{2\ln{2}} \gamma_\mathrm{gain} = \qty{20}{\meV}$ (Table\,\ref{table:simulation_parameters}). Indeed, the shape of the gain peak is predominantly determined by the in-plane dispersion. 
Lasing starts when the modal gain maximum matches the sum of the mirror and waveguide losses, 
\begin{equation}\label{eq:gain_match_loss}
	\Gamma g_\mathrm{th}
	=
	\frac{1}{L_\mathrm{cav}}\ln{\left(\frac{1}{R}\right)}
	+
	\alpha_\mathrm{w},
\end{equation}
where $L_\mathrm{cav}$, $R$, and $\Gamma$ are the cavity length, mirror reflectivity, and confinement factor of the optical mode in the cascade region, respectively. The waveguide loss $\alpha_\mathrm{w}$ is caused by the cladding and separate confinement layers. The voltage is varied until Eq.\,\eqref{eq:gain_match_loss} is satisfied (Fig.\,\ref{fig:flowchart}), defining the threshold voltage $V_\mathrm{th}$.

We analyze the threshold carrier balance in the W-QW by the average densities (\unit{\per\cm\cubed}),
\begin{subequations}\label{eq:definition_average_carrier_density}
	\begin{align}
		\bar{n}_\mathrm{th}^\mathrm{w}
		&\equiv
		\frac{1}{w_\mathrm{e}}
		\int_\mathrm{WQW} n_\mathrm{th}(z) \spaced{z}, \\
		\bar{p}_\mathrm{th}^\mathrm{w}
		&\equiv
		\frac{1}{w_\mathrm{h}}
		\int_\mathrm{WQW} p_\mathrm{th}(z) \spaced{z},
	\end{align}
\end{subequations}
where the spatial extent of the upper (lower) laser states, $w_\mathrm{e(h)}=\int_\mathrm{WQW} |\psi_\mathrm{e(h)}(z)|^2 \spaced{z} / |\psi_\mathrm{e(h)}|_\mathrm{max}^2 = 2.96\ (2.11)$~\unit{\nm}, remain approximately constant in this study as we do not modify the W-QW. 
Not normalizing by $w_\mathrm{e(h)}$, i.e., comparing the electron and hole sheet densities, would miss the significant difference in spatial confinement of the electrons and holes.
Similarly, the amount of carriers in the injectors are denoted by the volume densities $\bar{n}_\mathrm{th}^\mathrm{inj}$ and $\bar{p}_\mathrm{th}^\mathrm{inj}$~(\unit{\per\cm\cubed}).

\subsection{Auger recombination rates}
\label{subsec:Auger_rate}
Since the Auger mechanism dominates the non-radiative recombination of ICLs at room temperature, we assess different injector designs in terms of Auger rates. The rates are usually modeled phenomenologically and read in the non-degenerate (Maxwell--Boltzmann) limit~\cite{Blakemore2002semiconductor,Dutta1983}
\begin{align}
	R_\mathrm{A}
	&=
	C_\mathrm{n} n^2p
	+
	C_\mathrm{p} p^2n, \label{eq:Auger_rate_nondegenerate}
\end{align}
where $C_\mathrm{n/p}$ characterize the strength of the $eeh$ and $ehh$ Auger processes, respectively. The ratio $C_\mathrm{n} / C_\mathrm{p}$ would dictate how the carrier rebalancing reduces the threshold current, but has uncertainty due to the following reasons. 
First, $C_\mathrm{n/p}$ depend on temperature and device geometry as they reflect the availability of high-energy states for the \quotes{partner} carrier to be scattered into. Eq.\,\eqref{eq:Auger_rate_nondegenerate} can be fitted to experiments, but cannot predict the Auger recombination in newly-designed QWs. As the $eeh$ and $ehh$ processes cannot be measured individually, the estimations of ICL Auger coefficients~\cite{Bewley2008,Meyer2021} assumed equal volume densities of electrons and holes. While it is a good approximation for photodetectors and QW lasers, it obscures the relation between the carrier balance and Auger rates in ICLs. 
Second, Eq.\,\eqref{eq:Auger_rate_nondegenerate} grows as a cubic function in carrier densities, but measurements show a deviation from it in the degenerate regime, in which the probability of finding the recombining electron-hole pair approaches unity. Experiments and calculations for mid-infrared QW devices~\cite{Flatte1999,Hader2005_microscopic_theory,Hader2005} confirm this; the density dependence reduces to quadratic or even less in the degenerate regime. This makes the fitting of Eq.\,\eqref{eq:Auger_rate_nondegenerate} to experiments challenging. 
The non-degenerate approximation seems inappropriate for the present ICL structure, too. Figs.\,\ref{fig:ICL_states_densities}(a--b) show that, at the lasing threshold, the ground states in the conduction and valence bands overlap with the quasi-Fermi levels. 
As a result, Eq.\,\eqref{eq:Auger_rate_nondegenerate} is limited in explaining the correlation between the carrier balance and Auger rates in ICLs.

Microscopically, the Auger process and its inverse process, impact ionization, are types of carrier-carrier scattering via Coulomb interaction. The interaction is short-range due to the screening by other carriers. Earlier works \cite{Bockelmann1992,Gilard1998,Flatte1999,Grein2002_article,Olson2015_Auger} studied the scattering under the Born approximation. Calculation based on Fermi's golden rule does not rely on phenomenological parameters and is therefore predictive. The method has explained Auger rates in several QW designs~\cite{Grein2002_article} and at low temperatures~\cite{Olson2015_Auger}. 
Furthermore, Hader et al.~\cite{Hader2005_microscopic_theory,Hader2005} solved the semiconductor Bloch equations, including the carrier-carrier scattering in the second Born--Markov approximation. The calculated threshold currents of near-infrared QW lasers matched well with the measurements up to room temperature.

In view of future integration with quantum transport, we opt here for the NEGF formalism. Carrier-carrier scattering can be included at different levels of sophistication. Since the Hartree--Fock approximation neglects the inelastic scattering and hence the Auger process, we use the next-order treatment of the Coulomb interaction---the $GW$ approximation---described in Appendix~\ref{app:carrier_carrier_scattering}. It has been used to improve the accuracy of current-voltage characteristics of THz quantum cascade lasers~\cite{Kubis2009,Winge2016_GW} and nanowire field-effect transistors~\cite{Deuschle2025}. Its recent application to hot carriers in QW solar cells~\cite{Aeberhard2019} and Auger recombination in carbon nanotubes~\cite{Deuschle2023} and type-II superlattices~\cite{Montoya2025} gives us confidence that the method is sufficiently accurate for Auger recombination in ICLs.
Electrons are scattered into (out of) a state at energy $E$ at the rate given by the lesser (greater) components, $\bm{\Sigma}^{<(>)}(E)$, of the $GW$ self-energy.

The net interband current due to the carrier-carrier scattering is given by Auger rate minus impact ionization rate. In quasi-equilibrium, the ratio of the latter to former is $\exp{\left[- (E_\mathrm{F}^\mathrm{c} - E_\mathrm{F}^\mathrm{v}) / (\kBTe) \right]}$~\cite{Grein1995}, where $k_\mathrm{B}$ and $T_\mathrm{e}$ are the Boltzmann constant and the carrier temperature, respectively. In mid-infrared ICLs at room temperature close to threshold, $E_\mathrm{F}^\mathrm{c} - E_\mathrm{F}^\mathrm{v} \gtrsim E_\mathrm{ph} \gg \kBTe$, where $E_\mathrm{ph}$ is the photon energy, so that impact ionization is negligible.
The interband current due to the Auger mechanism is then given by
\begin{equation}\label{eq:GW_interband_current}
	J_\mathrm{A}
	=
	\frac{e}{\hbar} 
	\int \frac{\spaced{\mathbf{k}_\parallel}}{(2\pi)^2}
	\int_\mathrm{VB} \frac{\spaced{E}}{2\pi}
	\mathrm{Tr}\!
	\left[
		\bm{\Sigma}^<(\mathbf{k}_\parallel, E) 
		\bm{G}^>(\mathbf{k}_\parallel, E)
	\right],
\end{equation}
where $\bm{G}^>$ is the greater Green's function of electrons. The energy integration is performed below the heterostructure band gap.
In steady states, the lesser component of the $GW$ self-energy is represented in the RRS basis as (see Appendix~\ref{app:carrier_carrier_scattering})
\begin{align}\label{eq:self_energy_piecewise_GW}
    \Sigma_{\alpha\beta}^<(\mathbf{k}_\parallel, t)
    &=
    \imag
	\int \frac{\spaced{\mathbf{q}_\parallel}}{(2\pi)^2}
    G_{\alpha\beta}^<(\mathbf{k}_\parallel + \mathbf{q}_\parallel, t) \notag\\
	&\quad \times
	\left[
		W_{\alpha\beta}^{\mathrm{c}<}(\mathbf{q}_\parallel, t)
		+
		W_{\alpha\beta}^{\mathrm{v}<}(\mathbf{q}_\parallel, t)
	\right],
\end{align}
where $\bm{W}^{\mathrm{c/v}<}(t) = \bm{W}^{\mathrm{c/v}>}(-t)$ are the Fourier-transforms of
\begin{align}
	&W_{\alpha\beta}^{\mathrm{c/v}>}(\mathbf{q}_\parallel, E')
	\notag\\
	&\qquad
	=
	\sum_{\gamma\delta}
	W_{\alpha\gamma}^\mathrm{R}(\mathbf{q}_\parallel, E') 
	\Pi_{\gamma\delta}^{\mathrm{c/v}>}(\mathbf{q}_\parallel, E') 
	W_{\delta\beta}^\mathrm{A}(\mathbf{q}_\parallel, E'). \label{eq:W_gtr_energy_domain}
\end{align}
We distinguish $\Pi^\mathrm{c}$ and $\Pi^\mathrm{v}$ by the range of the energy integration; they describe the conduction and valence intersubband transitions, respectively, of the \quotes{partner} electron. The first and second terms in Eq.\,\eqref{eq:self_energy_piecewise_GW} therefore correspond to the $eeh$ and $ehh$ Auger processes, respectively.
The retarded and advanced screened Coulomb potentials between two particles in the RRS modes $\alpha$ and $\beta$ are (Appendix~\ref{app:carrier_carrier_scattering})
\begin{align}
	W_{\alpha\beta}^\mathrm{R/A}(\mathbf{q}_\parallel)
	\approx
	\frac{e^2}{2\epsilon_0\epsilon(0)L^2}
	\frac{\ee{-q_\mathrm{s} |z_\alpha - z_\beta|}}{q_\mathrm{s}}, \label{eq:W_RA_in_energy}
\end{align}
where $\epsilon(0)$ is the static dielectric constant averaged over one period, $L^2$ the normalization area, and $q_\mathrm{s} = \sqrt{\lambda_\mathrm{s}^{-2} + q_\parallel^2}$. A screening length $\lambda_\mathrm{s}$ of \qty{4}{\nm} is assumed to be reasonable based on the plasma screening models of Fermi gases (see Appendix~\ref{app:carrier_carrier_scattering}).
Since $\bm{W}^\mathrm{R/A}(\mathbf{q}_\parallel)$ decays exponentially in the reciprocal space, the $\mathbf{q}_\parallel$-dependence of $\bm{\Pi}^{>}$ in Eq.\,\eqref{eq:W_gtr_energy_domain} is irrelevant. We therefore approximate the polarization function by its $q_\parallel = 0$ value.
In general, the polarization function has two mode indices in the RRS basis (see Appendix~\ref{app:carrier_carrier_scattering}).
The scattered (hot) electrons in the excited subbands are expected to have short coherence lengths due to other scattering mechanisms not considered here. Therefore, we only consider the diagonal terms of the polarization functions,
\begin{align}
    \Pi_{\alpha\alpha}^{\mathrm{c}>}(0, E')
	&=
    -\imag
    \int \frac{\spaced{\mathbf{q}'_\parallel}}{(2\pi)^2}
    \int_{E_\mathrm{b}}^\infty 
	\frac{\spaced{E''}}{2\pi} \notag\\
	&\quad \times
    G_{\alpha\alpha}^>(q_\parallel', E' + E'') 
    G_{\alpha}^<(q_\parallel', E''), \label{eq:polarization_lessgtr_in_energy_c} \\
	\Pi_{\alpha\alpha}^{\mathrm{v}>}(0, E')
	&=
    -\imag
    \int \frac{\spaced{\mathbf{q}'_\parallel}}{(2\pi)^2}
    \int_{-\infty}^{E_\mathrm{b} - E'} 
	\frac{\spaced{E''}}{2\pi} \notag\\
	&\quad \times
    G_{\alpha\alpha}^>(q_\parallel', E' + E'') 
    G_{\alpha\alpha}^<(q_\parallel', E''). \label{eq:polarization_lessgtr_in_energy_v}
\end{align}
As long as the Auger recombination is slower than the intraband transport, i.e., quasi-equilibrium approximation is valid, the $GW$ self-energy needs to be evaluated just once after the charge distribution and electrostatic potential have converged (Fig.\,\ref{fig:flowchart}).

In the following, we assess by Eq.\,\eqref{eq:GW_interband_current} how the injector design affects the carrier balance and threshold current. 

\section{Results and Discussion}
\label{sec:results_and_discussion}

\begin{figure}
    \centering
	\ifFlatFolderStructure
		\includegraphics[width=\linewidth]{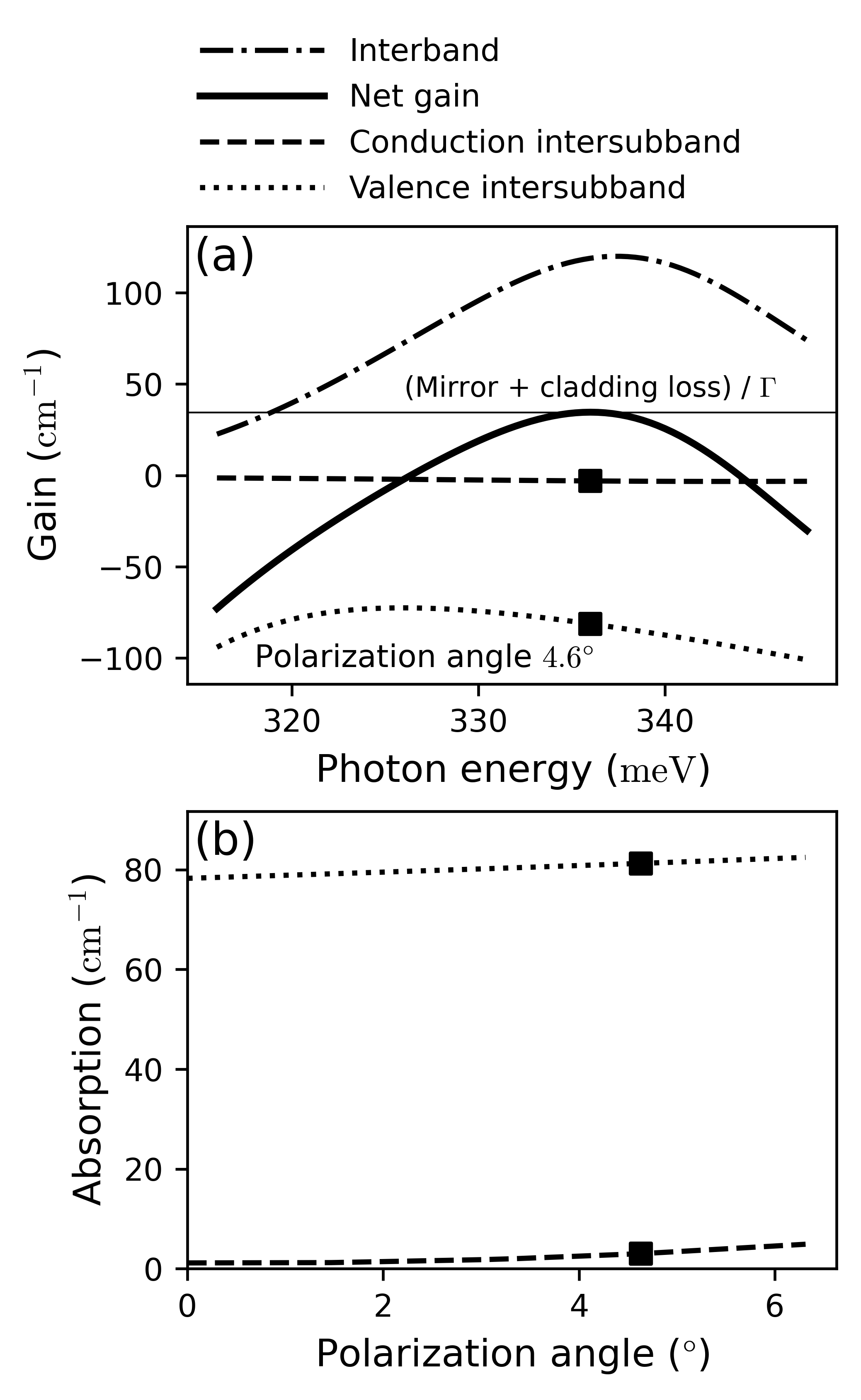}
	\else
		\includegraphics[width=\linewidth]{./images/paper/nextnanopy/absorption_vs_polarization_doping2D_4.76e11_P6_375.384mV_95TE_GW_NVBO_axial1000500.png}
	\fi
    \caption{
		Optical spectra of the reference design in Table~\ref{table:design_variations} at the threshold voltage $V_\mathrm{th} = \qty{375}{\mV}$.
		(a) Net gain spectrum and its decomposition into the interband, conduction intersubband and valence intersubband contributions. The thin solid line indicates the losses outside of the cascade stages. 
		(b) Parasitic absorption at gain peaks as a function of the polarization angle measured from the $x$--$y$ plane.
		In both panels, squares mark the values extracted for the analysis of the parasitic absorption in Fig.\,\ref{fig:comparison_doping_vs_In}.
	}
	\label{fig:gain_spectrum}
\end{figure}

\begin{figure*}
	\centering
	\ifFlatFolderStructure
		\includegraphics[width=\linewidth]{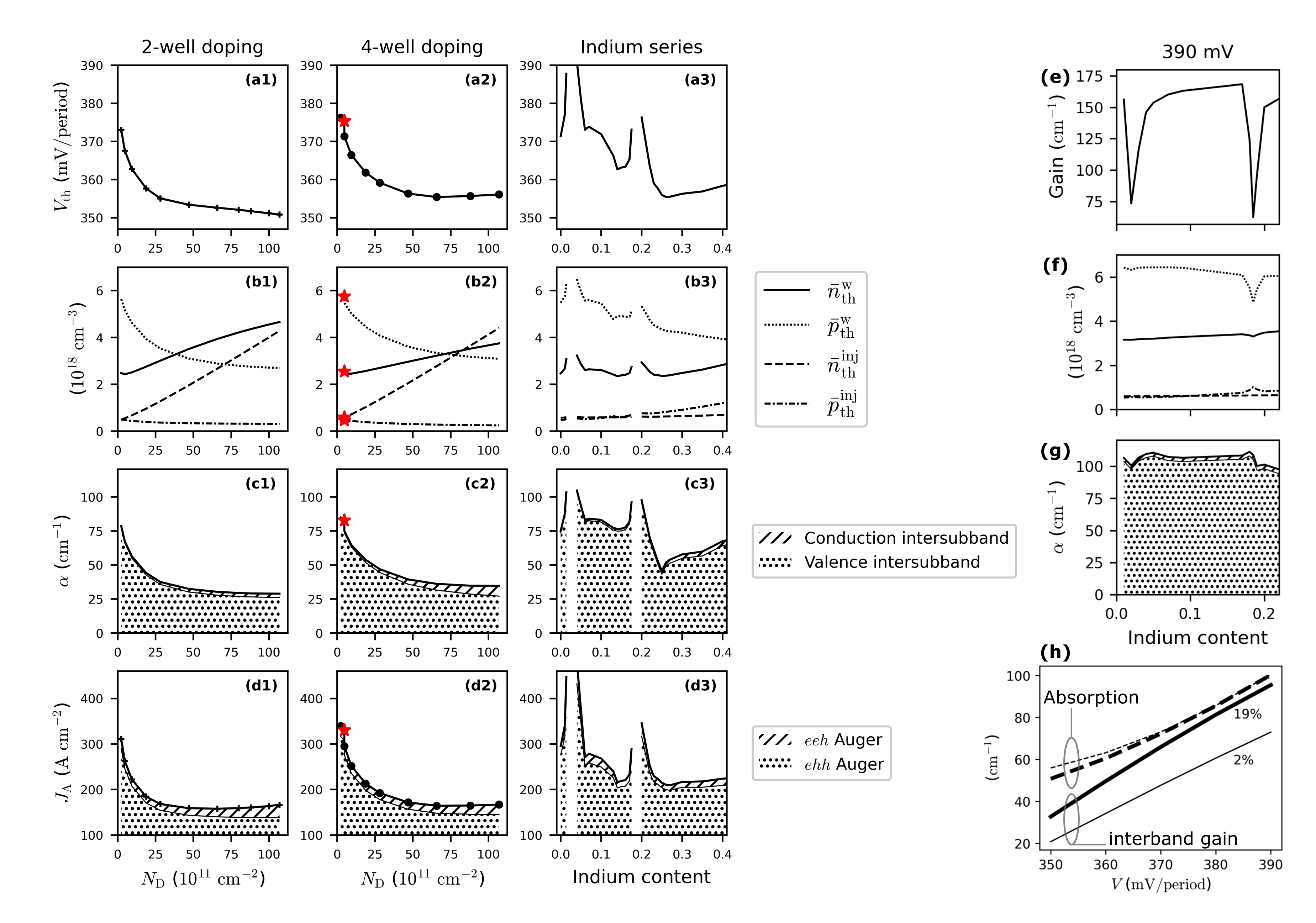} 
	\else
		\includegraphics[width=\linewidth]{./images/paper/comparison_doping_vs_In_390mV.png} 
	\fi
	\caption{
		Impact of the injector design variations on ICL performance figures. Red stars indicate the reference structure in Table~\ref{table:design_variations}.
		(a) Potential drop per period at the lasing thresholds.
		(b) Threshold carrier densities in the W-QW and injectors in the doping and indium series. The W-QW carrier densities are defined by Eqs.\,\eqref{eq:definition_average_carrier_density}, whereas the injector counterparts are simply averaged by the injector lengths.
		(c) Optical absorption coefficients at the photon energies of gain peaks and their breakdown to the conduction and valence intersubband contributions.
		(d) Interband currents due to the $eeh$ and $ehh$ Auger mechanisms.
		(e) Peak values of the gain without parasitic absorptions, (f) carrier densities, and (g) intersubband absorptions at the photon energy of net gain peaks from the indium series at a fixed bias of \qty{390}{\mV}.
		(h) Bias-dependence of the raw gain (solid) and parasitic absorption (dashed) at the indium contents of 2\% (thin) and 19\% (thick).
	}
	\label{fig:comparison_doping_vs_In}
\end{figure*}

We first investigate the gain spectra.
Fig.\,\ref{fig:gain_spectrum}(a) illustrates the net gain spectrum of the 4-well doping design with $y = \qty{5e18}{\per\cm\cubed}$, and its decomposition into the interband (raw) gain and parasitic absorptions.
The net gain peaks at the wavelength \qty{3.73}{\um}, which is within the experimentally observed range~\cite{Vurgaftman2011}.
In general, the raw gain is reduced by both conduction and valence intersubband absorptions because the \kp coupling blurs the selection rules. In addition, an oblique light polarization activates both optical transitions. Although ICL laser levels barely couple to TM-polarized light, the polarization component of the lasing-mode along the growth may not be negligible~\cite{HedwigPhD}. 
Fig.\,\ref{fig:gain_spectrum}(b) shows that the parasitic absorptions at the photon energy of the net gain peaks increase with the polarization angle measured from the $x$--$y$ plane.
In the absence of concrete data about the light polarization in ICLs, we choose a slightly oblique angle of \ang{4.6} to estimate the upper bound of the Auger rates. When we assume a TE polarization (\ang{0}), the threshold voltage and Auger currents we will present in Fig.\,\ref{fig:comparison_doping_vs_In} uniformly decrease.

Fig.\,\ref{fig:comparison_doping_vs_In} compares various properties of the designs in Table~\ref{table:design_variations}. The first three columns show the results at the threshold voltages in (a).
Figs.\,\ref{fig:comparison_doping_vs_In}(b1--b2) show that the electron densities in the W-QW and the electron injector increase approximately linearly with the sheet doping density. Heavily doping the electron injector raises both quasi-Fermi levels, thereby decreasing the hole concentration in the W-QW. Our model thus reproduces the carrier rebalancing in W-QWs~\cite{Vurgaftman2011}. 
Figs.\,\ref{fig:comparison_doping_vs_In}(c1--c2) track the parasitic absorptions at the net gain peak (i.e., the squares in Figs.\,\ref{fig:gain_spectrum}), showing that the absorption is much stronger in the valence band due to the light polarization almost perpendicular to the growth direction. The electron injector doping collectively reduces the threshold voltage [Figs.\,\ref{fig:comparison_doping_vs_In}(a1--a2)], the hole concentration [Figs.\,\ref{fig:comparison_doping_vs_In}(b1--b2)], and the valence intersubband absorption [Figs.\,\ref{fig:comparison_doping_vs_In}(c1--c2)]. The trend stagnates at higher doping densities. 

At a fixed sheet doping density, the 4-well doping reduces the band bending of the electron injector and populates the widest (i.e., first) QW more, thereby increasing the conduction intersubband absorption compared to the 2-well doping. Nevertheless, it remains a minor contribution to the total parasitic absorption in the cascade period. The heavy doping of the electron injector therefore improves carrier balance and reduces the valence intersubband absorption, with the effect being mostly insensitive to which electron injector wells are doped. 

Figs.\,\ref{fig:comparison_doping_vs_In}(d1--d2) depict the interband currents calculated by Eq.\,\eqref{eq:GW_interband_current}. The $eeh$ process is significantly slower than the $ehh$ one even at the perfect carrier balance in Figs.\,\ref{fig:comparison_doping_vs_In}(a1--a2). In the phenomenological model [Eq.\,\eqref{eq:Auger_rate_nondegenerate}], this would mean that $C_\mathrm{p} > C_\mathrm{n}$, but the validity of the model is questionable in the degenerate regime (see Sec.\,\ref{subsec:Auger_rate}).
The $eeh$ process is much slower because the subbands are sparser in the conduction than in the valence band ($\bm{\Pi}^\mathrm{c}$ is smaller than $\bm{\Pi}^\mathrm{v}$). While the $eeh$ contribution increases with doping, the $ehh$ rate drops more significantly due to the decreasing hole concentration in the W-QW, resulting in an overall reduction of the Auger current.
Our results show that reducing the hole population in the W-QW is effective for minimizing threshold currents owing to the selection rules and dense LDOS in the valence band.

The relative contribution of the $eeh$ mechanism to the total Auger current is larger in the 2-well than in the 4-well doping series in Figs.\,\ref{fig:comparison_doping_vs_In}(d1--d2) because the carrier rebalancing in the W-QW occurs at lower doping levels in Fig.\,\ref{fig:comparison_doping_vs_In}(b1).

\begin{figure}
    \centering
	\ifFlatFolderStructure
		\includegraphics[width=\linewidth]{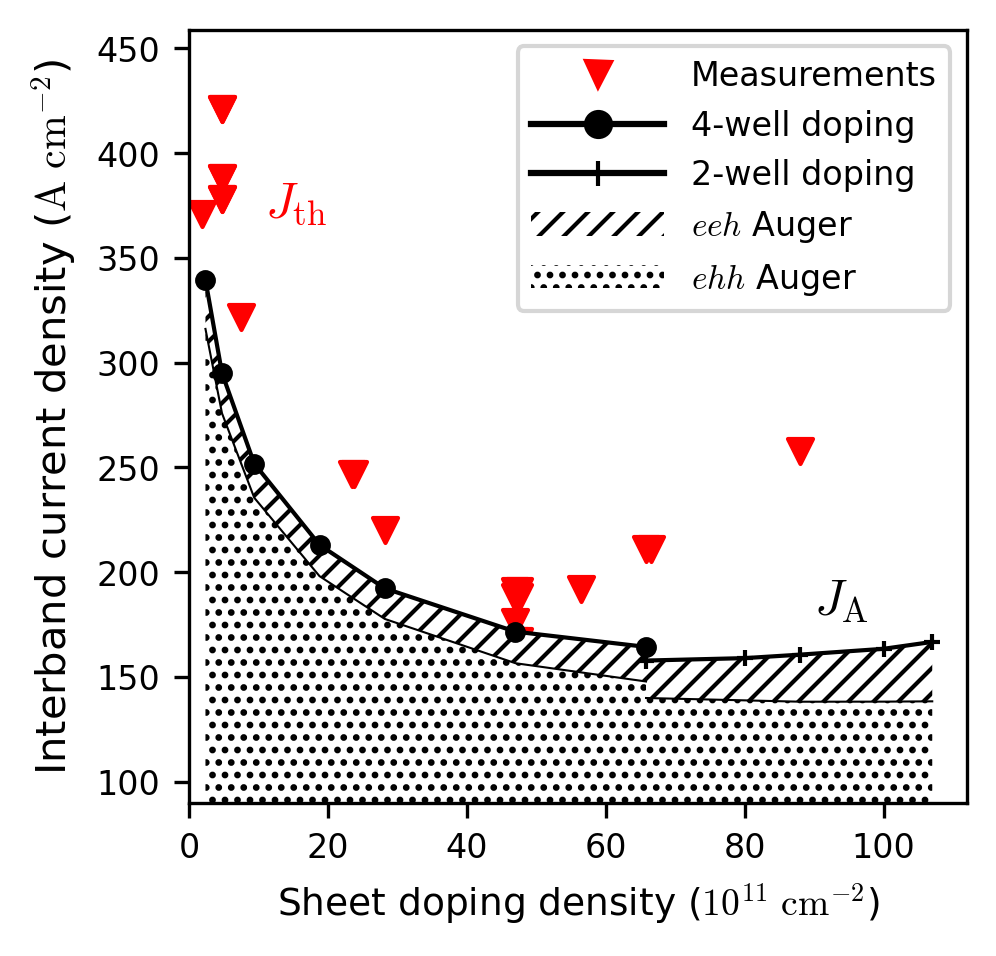}
	\else
		\includegraphics[width=\linewidth]{./images/paper/nextnanopy/GWJth_vs_sheetdoping.png}
	\fi
    \caption{
		Threshold current densities from the measurements~\cite{Vurgaftman2011,WeihPhD,VurgaftmanPrivateComm} (triangles) and Auger currents from the simulations (curves) against sheet doping density. 
	}
	\label{fig:GWJth_vs_sheetdoping}
\end{figure}

The strategy above was to reduce the hole population in the W-QW indirectly by doping the electron injector. We note in Figs.\,\ref{fig:comparison_doping_vs_In}(b1--b2) and (d1--d2) that most of the holes reside in the W-QW and contribute to the $ehh$ Auger process.
This motivates another strategy: distribute more holes to the injector. However, experimental observations discourage raising the subbands in the two hole injector QWs closer to or higher than the topmost hole state in the W-QW~\cite{Meyer2020}. To elucidate the reason for this empirical rule of thumb, we introduce indium to the hole injector QWs. The sheet doping density of the electron injector is set equal to the reference structure (see Table~\ref{table:design_variations}). Experimentally, the indium content in such material combination is commonly kept not exceeding 40\% to avoid high strain and defects. The hole injector levels rise by \qtyrange{85}{90}{\meV} when the indium content is increased from 0\% to 40\%. We checked that the cutoff energy $E_\mathrm{cutoff}^\mathrm{v}$ in Table~\ref{table:simulation_parameters} was sufficiently large to include all the confined states in the valence band in the RRS basis construction throughout this alloy content range.
As the threshold was not reached around 2\% and 19\%, corresponding data are omitted in Figs.\,\ref{fig:comparison_doping_vs_In}(a3--d3).

Fig.\,\ref{fig:comparison_doping_vs_In}(b3) shows the threshold carrier densities in the indium series and confirms the carrier rebalancing in the W-QW, but the effect is weaker than in the injector doping series. 
In contrast to Figs.\,\ref{fig:comparison_doping_vs_In}(b1--b2), the indium series barely increases the electron population in the first electron injector QW, which suppresses the conduction intersubband absorption in Fig.\,\ref{fig:comparison_doping_vs_In}(c3). The benefit is smaller, however, than the cost of the higher hole concentration in the hole injector, which enhances the parasitic valence intersubband absorption.

The sharp spikes in the indium series can be understood from the behavior at a fixed bias voltage of 390~$\mathrm{mV}$. 
Fig.\,\ref{fig:comparison_doping_vs_In}(e) tracks the raw gain at the net gain peak in Fig.\,\ref{fig:gain_spectrum}(a). 
Figs.\,\ref{fig:comparison_doping_vs_In}(f--g) show the carrier densities and parasitic absorptions at the bias, respectively.
The raw gain exhibits two dips near the indium contents of 2\% and 19\%, corresponding to the anomalies in Fig.\,\ref{fig:comparison_doping_vs_In}(a3). The second dip is accompanied by a hole redistribution from the W-QW to the hole injector at 19\% [Fig.\,\ref{fig:comparison_doping_vs_In}(f)].
By studying the wavefunctions, we attribute the two dips in Fig.\,\ref{fig:comparison_doping_vs_In}(e) to a hybridization of the hole states in the W-QW and hole injector. 
In contrast, Fig.\,\ref{fig:comparison_doping_vs_In}(g) shows that the parasitic absorption is less sensitive to the indium content. 
Fig.\,\ref{fig:comparison_doping_vs_In}(h) shows how the raw gain and parasitic absorption at the net gain peak grow with the bias. The gain still grows slightly faster than the parasitic absorptions. However, the degraded raw gain prevents the laser from reaching the threshold.

The threshold voltage appears compelling at the indium content of 25\% to 40\% compared with the doping series. However, Figs.\,\ref{fig:comparison_doping_vs_In}(a3) and (d3) show that a higher hole density in the W-QW increases the $ehh$ Auger recombination. This underlines again the importance of reducing the hole concentration for minimizing threshold currents. To summarize, the carrier rebalancing strategy by means of raising the hole injector levels fails to efficiently reduce the hole concentration in the recombination QWs and suffers from a stronger Auger recombination.

Fig.\,\ref{fig:GWJth_vs_sheetdoping} compares the Auger recombination currents calculated by Eq.\,\eqref{eq:GW_interband_current} with the measurements~\cite{Vurgaftman2011,VurgaftmanPrivateComm,WeihPhD}.
As the $ehh$ process is dominant, reducing the hole population in the W-QW significantly decreases the threshold current in the left half of the plot.
While the measured threshold currents increase in the right half, our simulation predicts a flatter trend. The choice of $\lambda_\mathrm{s}$, $\gamma_\mathrm{gain}$, or the polarization vector shifts the curve up- or downward, but does not alter its trend. For example, the dotted curve in Fig.\,\ref{fig:GWJth_vs_sheetdoping_vs_intrasubband} in Appendix~\ref{app:sim_parameters} shows that a smaller value of $\gamma_\mathrm{gain}$ increases the peak gain, thereby reducing the threshold carrier densities and Auger rates. However, the qualitative trend remains similar.
At high doping, one might expect an enhanced parasitic absorption and hence a higher threshold voltage. However, the absorption is small in our calculation, as shown in Figs.\,\ref{fig:comparison_doping_vs_In}(c1--c2), due to the selection rules of the conduction intersubband transitions for nearly TE-polarized light.
Radiative recombination does not explain the increasing threshold current since our calculation (not shown) indicates that the spontaneous emission generates interband currents of merely \qtyrange{6}{9}{\ampere\per\cm\squared}. 
We do not expect a significant contribution from the Shockley--Read--Hall (SRH) recombination either since \qtyrange{3}{5}{\um} ICLs with type-II recombination QWs are known to be tolerant to dislocations~\cite{Cerutti2021,Fagot2025}.

One possible reason is the carrier escape from the W-QW. Experiments with ICLs grown on InAs or GaSb substrates imply that about 10\% to 20\% of the injected carriers either leak through sidewalls, or enter the laser levels but escape without recombination~\cite{Meyer2020}. The former does not depend on the injector doping. The latter mechanism may be activated at lasing thresholds by the Auger process; the carrier-carrier scattering promotes the excess carriers in the W-QW to higher subbands, helping them overcome the barriers and travel to the next period without recombination [see Fig.\,\ref{fig:ICL_states_densities}(b)]. This Auger-mediated carrier escape will depend on the doping level via $\bar{n}_\mathrm{th}^\mathrm{w}$ shown in Figs.\,\ref{fig:comparison_doping_vs_In}(b1) and (b2). If we attribute the discrepancy at $N_\mathrm{D} = \qty{8.8e12}{\per\cm\squared}$ in Fig.\,\ref{fig:GWJth_vs_sheetdoping} exclusively to this mechanism, about 36\% of the injected carriers escape from the W-QW. This exceeds the anticipated range from the measurement.

The second possible reason is \textit{intrasubband} absorptions mediated by scattering due to e.g.\ interface roughness~\cite{Dumke1961_FCA,Vurgaftman1999_absorption}. This mechanism may trigger the parasitic absorption particularly in the electron injector, which contains more and more electrons as we dope [Fig.\,\ref{fig:comparison_doping_vs_In}(b2)]. Our supplementary calculation in Appendix~\ref{app:sim_parameters} suggests that this absorption may raise $J_\mathrm{A}$ at higher doping concentrations.

\section{Conclusion and outlook}
\label{sec:conclusion}
We have studied the impact of ICL injector designs on threshold current density by calculating the threshold carrier densities and Auger recombination rates. The model combines electrostatic effects, interband coupling, and in-plane momentum- and polarization-dependent gain calculation with a microscopic calculation of the Auger rates.
Our findings are threefold. 
First, heavily doping the electron injector~\cite{Vurgaftman2011,WeihPhD} reduces the threshold current density essentially because the decrease of the hole concentration suppresses the dominant parasitic absorption and $ehh$-Auger process.
Second, in the absence of scattering-mediated carrier escape and light absorption, threshold currents do not increase at high doping densities. Such second-order contributions may explain the increase in the measurements.
Third, an alternative carrier rebalancing approach---raising the hole levels in the injector by $\mathrm{Ga}_{1-x}\mathrm{In}_x\mathrm{Sb}$ QWs---does not outperform the doping strategy. The indium series is less efficient in transferring holes from the W-QW to the hole injector, thereby suffering from a larger valence intersubband absorption at this emission wavelength. This necessitates a larger population inversion, resulting in faster Auger recombination.
Thus, our modeling of ICLs based on the 8-band \kp model and quasi-equilibrium Green's functions provides physical insights to the impact of carrier injector designs on lasing thresholds.

Throughout this study, we kept the layer thicknesses and number of wells of the injectors constant to focus on the effect of the doping profile and well depth. When these parameters are optimized accordingly for each target emission wavelength, Auger rates may be reduced further.

Our model offers a basis for future non-equilibrium quantum transport simulations beyond the quasi-equilibrium approximation used here. This may reveal insights into remaining questions such as (i) the limit of the quasi-equilibrium approximation, and (ii) design-dependence of the scattering-mediated carrier escape and light absorption.

\begin{acknowledgments}
	We thank European Union's Horizon 2020 research and innovation program under the Marie Sk\l{}odowska-Curie grant agreement no.\,956548 (QUANTIMONY) for financial support.
\end{acknowledgments}

\section*{Author Declarations}
\subsection*{Conflict of Interest}
The authors have no conflicts to disclose.

\subsection*{Author Contributions}
\textbf{Takuma Sato}: Conceptualization (equal); Data curation (lead); Formal analysis (equal); Methodology (equal); Software (lead); Visualization (lead); Writing - original draft (lead); Writing - review \& editing (equal).
\textbf{Borislav Petrovi\'{c}}: Conceptualization (equal); Data curation (supporting); Formal analysis (equal); Visualization (supporting); Writing - original draft (supporting); Writing - review \& editing (equal).
\textbf{Robert Weih}: Validation (equal); Writing - review \& editing (equal).
\textbf{Fabian Hartmann}: Validation (equal); Writing - review \& editing (equal).
\textbf{Sven H\"{o}fling}: Funding acquisition (equal); Validation (equal); Writing - review \& editing (equal).
\textbf{Stefan Birner}: Funding acquisition (equal); Project administration (lead); Resources (equal); Supervision (supporting); Writing - review \& editing (equal).
\textbf{Christian Jirauschek}: Resources (equal); Supervision (supporting); Validation (supporting); Writing - review \& editing (equal).
\textbf{Thomas Grange}: Conceptualization (equal); Methodology (equal); Software (supporting); Supervision (lead); Validation (equal); Writing - review \& editing (equal).

\section*{Data Availability}
The data that support the findings of this study are openly available in Zenodo at https://doi.org/10.5281/zenodo.17642241.

\appendix
\section{8-band \kp Hamiltonian}
\label{app:Hamiltonian}
We find the energy eigenstates in a heterostructure of zincblende crystals within the envelope function approximation. Each material layer is described by the following 8-band \kp Hamiltonian,
\begin{equation}\label{eq:8kp_Hamiltonian}
	\bm{H}_0(\mathbf{k}, \varepsilon)
	=
	\bm{H}_\mathrm{kp}^{8\times8}
	+
	\bm{H}_\Delta^{8\times8}
	+
	\bm{H}_\varepsilon^{8\times8},
\end{equation}
in the basis $\{ |S\uparrow\rangle, |X\uparrow\rangle, |Y\uparrow\rangle, |Z\uparrow\rangle, |S\downarrow\rangle, |X\downarrow\rangle, |Y\downarrow\rangle, |Z\downarrow\rangle \}$.
The \kp Hamiltonian,
\begin{equation}
	\bm{H}_\mathrm{kp}^{8\times8}
	=
	\begin{pmatrix}
		H_\mathrm{cc} & \bm{H}_\mathrm{cv} & & \\
		\bm{H}_\mathrm{vc} & \bm{H}_\mathrm{vv} & & \\
		& & H_\mathrm{cc} & \bm{H}_\mathrm{cv} \\
		& & \bm{H}_\mathrm{vc} & \bm{H}_\mathrm{vv}
	\end{pmatrix},
\end{equation}
contains the free kinetic energy $\hbar^2k^2/(2m_0)$, \kp interband coupling $\mathcal{O}(k)$, and the lowest order \kp coupling via remote bands $\mathcal{O}(k^2)$:
\begin{widetext}
	\begin{subequations}
		\begin{align}
			H_\mathrm{cc} 
			&=
			E_\mathrm{c}(T_\mathrm{L})
			+
			\frac{\hbar^2}{2m_0}
			\left(
				k_xSk_x + k_ySk_y + k_zSk_z
			\right), \label{H_cc}\\
			\bm{H}_\mathrm{cv} 
			&=
			\begin{pmatrix}
				\imag Pk_x + k_y\frac{B}{2}k_z + k_z\frac{B}{2}k_y 
				& \imag Pk_y + k_z\frac{B}{2}k_x + k_x\frac{B}{2}k_z 
				& \imag Pk_z + k_x\frac{B}{2}k_y + k_y\frac{B}{2}k_x
			\end{pmatrix}, \label{H_cv}\\
			\bm{H}_\mathrm{vc} 
			&=
			\begin{pmatrix}
				-\imag k_xP + k_z\frac{B}{2}k_y + k_y\frac{B}{2}k_z 
				& -\imag k_yP + k_x\frac{B}{2}k_z + k_z\frac{B}{2}k_x 
				& -\imag k_zP + k_y\frac{B}{2}k_x + k_x\frac{B}{2}k_y
			\end{pmatrix}^t, \label{H_vc}\\
			\bm{H}_\mathrm{vv} 
			&=
			\left( E_\mathrm{v,av} + \frac{\hbar^2k^2}{2m_0} \right)\bm{I}_{3} \notag\\
			+
			&\begin{pmatrix}
				k_xL'k_x + k_yMk_y + k_zMk_z & k_xN^{+'}k_y + k_yN^{-'}k_x & k_xN^{+'}k_z + k_zN^{-'}k_x \\
				k_yN^{+'}k_x + k_xN^{-'}k_y & k_xMk_x + k_yL'k_y + k_zMk_z & k_yN^{+'}k_z + k_zN^{-'}k_y \\
				k_zN^{+'}k_x + k_xN^{-'}k_z & k_zN^{+'}k_y + k_yN^{-'}k_z & k_xMk_x + k_yMk_y + k_zL'k_z
			\end{pmatrix}, \label{eq:H_vv}
		\end{align}
	\end{subequations}
\end{widetext}
where $m_0$ is the bare electron mass. $E_\mathrm{v,av}$ is the average of the heavy-hole, light-hole and spin-orbit split-off band offsets of bulk in the absence of strain. The conduction band offset is calculated from the valence band offset, spin-orbit splitting energy $\Delta_\mathrm{so}$, \qty{0}{\kelvin} band gap, and the Varshni shift as
\begin{align}
	E_\mathrm{c}(T_\mathrm{L}) 
	&= 
	E_\mathrm{v,av} + \Delta_\mathrm{so}/3 + E_\mathrm{g}(T_\mathrm{L}), \\
	E_\mathrm{g}(T_\mathrm{L}) 
	&= 
	E_\mathrm{g}(0) - \frac{\alpha_\mathrm{g} T_\mathrm{L}^2}{T_\mathrm{L} + \beta_\mathrm{g}}.
\end{align}
\begin{table*}[]  %[tb]
	\caption{
		Material parameters used in the simulations.
	}
	\label{table:material_parameters} 
	\begin{ruledtabular}
		\begin{tabular}{lddddll}
			Symbol               & \text{InAs}   & \text{AlSb}  & \text{GaSb}   & \text{InSb}   & Units                 & Ref.                          \\\hline
			$a$                  & 0.605\,83     & 0.613\,55    & 0.609\,59     & 0.647\,94     & \unit{\nm}            & \cite{VurgaftmanBandsPhotons} \\
			$C_{11}$             & 83.29         & 87.69        & 88.42         & 68.47         & \unit{\GPa}           & \cite{VurgaftmanBandsPhotons} \\
			$C_{12}$             & 45.26         & 43.41        & 40.26         & 37.35         & \unit{\GPa}           & \cite{VurgaftmanBandsPhotons} \\
			$C_{44}$             & 39.59         & 40.76        & 43.22         & 31.11         & \unit{\GPa}           & \cite{VurgaftmanBandsPhotons} \\
			$E_\mathrm{v,av}$    & -0.742\,3     & -0.635\,3    & -0.283\,3     & -0.27         & \unit{\eV}            & \cite{VurgaftmanBandsPhotons} \\
			$E_\mathrm{g}$       & 0.417         & 2.386        & 0.812         & 0.235         & \unit{\eV}            & \cite{VurgaftmanBandsPhotons} \\
			$\alpha_\mathrm{g}$  & 3.07\times10^{-4} & 4.2\times10^{-4} & 4.17\times10^{-4} & 3.20\times10^{-4} & \unit{\eV\per\kelvin} & \cite{VurgaftmanBandsPhotons} \\
			$\beta_\mathrm{g}$   & 191           & 140          & 140           & 170           & \unit{\kelvin}        & \cite{VurgaftmanBandsPhotons} \\
			$S$                  & -1            & -0.12        & -1            & 0.54          & -                     & \cite{VurgaftmanBandsPhotons} \\
			$E_\mathrm{P}$       & 19.5          & 18.7         & 25.8          & 23.3          & \unit{\eV}            & \cite{VurgaftmanBandsPhotons} \\
			$B$                  & 3.596         & 0            & 13.079        & 2.703         & \unit{\eV}            & \cite{Cartoixa_spintronics}   \\
			$\gamma_1$           & 19.7          & 5.18         & 13.4          & 34.8          & -                     & \cite{VurgaftmanBandsPhotons} \\
			$\gamma_2$           & 8.4           & 1.19         & 4.7           & 15.5          & -                     & \cite{VurgaftmanBandsPhotons} \\
			$\gamma_3$           & 9.3           & 1.97         & 6             & 16.5          & -                     & \cite{VurgaftmanBandsPhotons} \\
			$\kappa$             & 7.68          & 0.31         & 3.18          & 14.76         & -                     & \cite{Lawaetz1971}            \\
			$\Delta_\mathrm{so}$ & 0.367         & 0.676        & 0.76          & 0.81          & \unit{\eV}            & \cite{VurgaftmanBandsPhotons} \\
			$\Xi$                & -5.08         & -4.5         & -7.5          & -6.94         & \unit{\eV}            & \cite{VurgaftmanBandsPhotons} \\
			$a_\mathrm{v}$       & 1             & 1.4          & 0.8           & 0.36          & \unit{\eV}            & \cite{VurgaftmanBandsPhotons} \\
			$b$                  & -1.8          & -1.35        & -2            & -2            & \unit{\eV}            & \cite{VurgaftmanBandsPhotons} \\
			$d$                  & -3.6          & -4.3         & -4.7          & -4.7          & \unit{\eV}            & \cite{VurgaftmanBandsPhotons} \\
			$\epsilon(0)$        & 15.15         & 12.04        & 15.69         & 17.5          & -                     & \cite{LandoltBornstein}
		\end{tabular}
	\end{ruledtabular}
\end{table*}
\begin{table}[]  %[tb]
	\caption{
		Ternary bowing parameters used in the simulations.
	}
	\label{table:material_parameters_bow} 
	\begin{ruledtabular}
		\begin{tabular}{llll}
			Symbol               & $\mathrm{Ga}_{1-x}\mathrm{In}_{x}\mathrm{Sb}$ & Units                 & Ref.                          \\\hline
			$E_\mathrm{v,av}$    & -0.033                & \unit{\eV}            & \cite{VurgaftmanBandsPhotons} \\
			$E_\mathrm{g}$       & 0.415                 & \unit{\eV}            & \cite{VurgaftmanBandsPhotons} \\
			$\alpha_\mathrm{g}$  & 0                     & \unit{\eV\per\kelvin} & \cite{VurgaftmanBandsPhotons} \\
			$\beta_\mathrm{g}$   & 0                     & \unit{\kelvin}        & \cite{VurgaftmanBandsPhotons} \\
			$S$                  & 0                     & -                     & \cite{VurgaftmanBandsPhotons} \\
			$E_\mathrm{P}$       & 7.8                   & \unit{\eV}            & \cite{VurgaftmanBandsPhotons} \\
			$\Delta_\mathrm{so}$ & 0.1                   & \unit{\eV}            & \cite{VurgaftmanBandsPhotons}
		\end{tabular}
	\end{ruledtabular}
\end{table}
The dimensionless parameter $S$ contains the kinetic energy of free electron and remote-band contributions,
\begin{equation}\label{eq:definition_S}
	S
	=
	1
	+
	\frac{2}{m_0}
	\sum_{r\in\text{remote}}
	\frac{| \langle S\sigma| \hat{p}_x |r\sigma \rangle|^2}{E_\mathrm{c}(0) - E_r},
\end{equation}
which is independent of the spin $\sigma=\ \uparrow,\downarrow$.
The parameter $P = (\imag\hbar / m_0) \langle S\sigma | \hat{p}_x | X\sigma\rangle = (\imag\hbar / m_0) \langle S\sigma | \hat{p}_y | Y\sigma\rangle = (\imag\hbar / m_0) \langle S\sigma | \hat{p}_z | Z\sigma\rangle = \hbar\sqrt{E_\mathrm{P} / (2m_0)}$ couples the conduction and valence band states and is calculated from the Kane energy $E_\mathrm{P}$. The asymmetric ordering of the linear-$k$ terms is to guarantee that integration of the envelope-function \schroedinger equation across material interfaces yields the correct number of boundary conditions in the limit $S\to 0$~\cite{Foreman1997}.
This limit avoids spurious solutions originating from the highest-order $k$-term of the secular equation~\cite{Foreman1997,VurgaftmanBandsPhotons}. 
In order not to damage the band structure, we adjust $E_\mathrm{P}$ to preserve the conduction band effective mass of each layer,
\begin{equation}
	\frac{m_0}{m_\mathrm{e}(T_\mathrm{L})}
	=
	S 
	+ 
	\frac{E_\mathrm{P}}{3}
	\left[
		\frac{2}{E_\mathrm{g}(T_\mathrm{L})}
		+
		\frac{1}{E_\mathrm{g}(T_\mathrm{L}) + \Delta_\mathrm{so}}
	\right],
\end{equation}
when setting $S$ to 0.
$B$ is an additional interband coupling via remote bands and causes spin-splitting in crystals without bulk inversion symmetry \cite{FundamentalsOfSemiconductors}. Unlike the $P$ terms, we symmetrize the $B$ terms so that the bulk Hamiltonian is invariant under a $90^\circ$ rotation around $x$, $y$, and $z$ crystal axes. 
$L', M,$ and 
\begin{equation}
	N^{\pm'}
	=
	\frac{N'}{2} \mp \frac{\hbar^2}{2m_0}(3\kappa' + 1),
\end{equation}
are the valence-band equivalent of the $S$ parameter and have the dimension of $\hbar^2/(2m_0)$.
$L'$, $N'$, and $\kappa'$ are shifted from their 6-band counterparts since the contribution of the conduction band, which is treated exactly and not as a remote band in the 8-band model, must be subtracted off~\cite{BirnerPhD},
\begin{subequations}
	\begin{align}
		L'
		&\equiv
		L_6 + \frac{P^2}{E_\mathrm{g}(0)}, \\
		N'
		&=
		N_6 + \frac{P^2}{E_\mathrm{g}(0)}, \\
		\kappa'
		&=
		\kappa_6
		-\frac{1}{6}\frac{E_\mathrm{P}}{E_\mathrm{g}(0)},
	\end{align}
\end{subequations}
where the band gap is without the Varshni shift since the 6- and 8-band \kp theories are constructed at \qty{0}{\kelvin}.
The 6-band Dresselhaus--Kip--Kittel parameters are calculated from the Luttinger parameters as~\cite{Bahder1990}
\begin{subequations}
	\begin{align}
		L_6
		&=
		-\frac{\hbar^2}{2m_0}\left(\gamma_1 + 4\gamma_2 + 1\right), \\
		M
		&=
		-\frac{\hbar^2}{2m_0}\left(\gamma_1 - 2\gamma_2 + 1\right), \\
		N_6
		&=
		-\frac{\hbar^2}{2m_0} 6\gamma_3.
	\end{align}
\end{subequations}
The spin-orbit splitting of the valence band states is
\begin{equation}
	\bm{H}_{\Delta}^{8\times8} = 
    \begin{pmatrix}
        \bm{H}_{\Delta\uparrow\uparrow} & \bm{H}_{\Delta\uparrow\downarrow} \\
        \bm{H}_{\Delta\uparrow\downarrow}^\dagger & \bm{H}_{\Delta\uparrow\uparrow}^*
    \end{pmatrix},
\end{equation}
where
\begin{subequations}
	\begin{align}
		\bm{H}_{\Delta\uparrow\uparrow} 
		&= 
		\frac{\Delta_\mathrm{so}}{3}
		\begin{pmatrix}
			0 & 0 & 0 & 0 \\
			0 & 0 & -\imag & 0  \\
			0 & \imag & 0 & 0 \\
			0 & 0 & 0 & 0
		\end{pmatrix},  \\
		\bm{H}_{\Delta\uparrow\downarrow}
		&= 
		\frac{\Delta_\mathrm{so}}{3}
		\begin{pmatrix}
			0 & 0 & 0 & 0 \\
			0 & 0 & 0 & 1 \\
			0 & 0 & 0 & -\imag \\
			0 & -1 & \imag & 0
		\end{pmatrix}.
	\end{align}
\end{subequations}

The elastic deformation energy is considered to first order in the strain tensor field $\bm{\varepsilon}(z)$,
\begin{equation}
	\bm{H}_\varepsilon^{8\times8}
	=
	\begin{pmatrix}
		D_\mathrm{c}(\varepsilon) & & & \\
		& \bm{D}_\mathrm{v}(\varepsilon) & & \\
		& & D_\mathrm{c}(\varepsilon) & \\
		& & & \bm{D}_\mathrm{v}(\varepsilon)
	\end{pmatrix},
\end{equation}
where
\begin{widetext}
	\begin{subequations}
		\begin{align}
			D_\mathrm{c}(\varepsilon) 
			&=
			\Xi(\varepsilon_{11}+\varepsilon_{22} + \varepsilon_{33}), \label{strainHamWZ_CB}\\
			\bm{D}_\mathrm{v}(\varepsilon) 
			&=
			\begin{pmatrix}
				l_\varepsilon \varepsilon_{11} + m_\varepsilon (\varepsilon_{22}+\varepsilon_{33}) & n_\varepsilon \varepsilon_{12} & n_\varepsilon \varepsilon_{13} \\
				n_\varepsilon \varepsilon_{12} & l_\varepsilon \varepsilon_{22} + m_\varepsilon (\varepsilon_{11}+\varepsilon_{33}) & n_\varepsilon \varepsilon_{23} \\
				n_\varepsilon \varepsilon_{13} & n_\varepsilon \varepsilon_{23} & l_\varepsilon \varepsilon_{33} + m_\varepsilon (\varepsilon_{11}+\varepsilon_{22})
			\end{pmatrix}. \label{strainHamWZ_VB}
		\end{align}
	\end{subequations}
\end{widetext}
$\bm{\varepsilon}(z)$ is calculated for each heterostructure layer from its lattice constant $a$ and the elastic stiffness constants $C_{ij}$ assuming pseudomorphic growth on a GaSb substrate. The conduction band is only sensitive to hydrostatic strains (coefficient $\Xi$) as the band has spherical symmetry. The coefficients in the valence band consist of the hydrostatic, uniaxial, and shear deformation potentials $(a_\mathrm{v}, b, d)$: $l_\varepsilon = a_\mathrm{v} + 2b, m_\varepsilon = a_\mathrm{v} - b,$ and $n_\varepsilon = \sqrt{3}d$.
All the material parameters are listed in Tables~\ref{table:material_parameters} and \ref{table:material_parameters_bow}.

The heterostructure Hamiltonian is a stack of the Hamiltonian \ref{eq:8kp_Hamiltonian} with $k_z$ replaced with the differential operators due to translational asymmetry along the growth direction,
\begin{equation}\label{eq:8kp_Hamiltonian_hetero}
    \bm{H}
	=
	\bm{H}^{(0)}
    + \bm{H}^{(L)} \frac{\mathrm{d}}{\spaced{z}}
    + \frac{\mathrm{d}}{\spaced{z}} \bm{H}^{(R)}
    + \frac{\mathrm{d}}{\spaced{z}} \bm{H}^{(2)} \frac{\mathrm{d}}{\spaced{z}},
\end{equation}
where $\bm{H}^{(0)}, \bm{H}^{(L)}, \bm{H}^{(R)}$, and $\bm{H}^{(2)}$ are $8\times8$ matrices dependent on position. The 8-band Hamiltonian is discretized on a real-space numerical grid by the finite difference method.  
We denote the resulting matrix by $\bm{H}^{8N \times 8N}$.
Our \kp solver is resistant to spurious solutions by virtue of the rescaling of $S$ described above and a consistent discretization of the differential operators. The first derivatives in Eq.\,\eqref{eq:8kp_Hamiltonian_hetero} are discretized with forward and backward rather than centered differences to avoid oscillatory solutions~\cite{HackenbuchnerPhD,Frensley2015_discretization}.
The Hermiticity of the resulting matrix and the consistency in the allocation of material parameters~\cite{Frensley2015_discretization} restrict the choice of discretization schemes for the differential operators in Eq.\,\eqref{eq:8kp_Hamiltonian_hetero}.

\section{Simulation parameters}
\label{app:sim_parameters}

\begin{table}[]  %[tb]
	\caption{
		Geometry and numerical parameters used in the simulations.
	}
	\label{table:simulation_parameters} 
	\begin{ruledtabular}
		\begin{tabular}{lrl}
			Symbol & Value & Units  \\\hline
			$T_\mathrm{L}$ & 300 & \unit{\kelvin} \\
			$T_\mathrm{e}$ & 300 & \unit{\kelvin} \\
			$E_\mathrm{cutoff}^\mathrm{c}$ & 1000 & \unit{\meV} \\
			$E_\mathrm{cutoff}^\mathrm{v}$ & 500 & \unit{\meV}  \\
			$E_\mathrm{cutoff}^\parallel$ & 300 & \unit{\meV}  \\
			$\gamma_\mathrm{subband}$ & 15 & \unit{\meV}  \\
			$\gamma_\mathrm{gain}$ & $20 / 2\sqrt{2\ln{2}}$ & \unit{\meV} \\
			$\polarizationVector$ & $(0.997, 0, 0.081)$ & -  \\
			$L_\mathrm{cav}$ & 2 & \unit{\mm} \\
			$R$ & 0.31 & - \\
			$\Gamma$ & 0.217 & - \\
			$\alpha_\mathrm{w}$ & 1.63 & \unit{\per\cm} \\
			$\lambda_\mathrm{s}$ & 4 & \unit{\nm}
		\end{tabular}
	\end{ruledtabular}
\end{table}

\begin{figure}
    \centering
	\ifFlatFolderStructure
		\includegraphics[width=\linewidth]{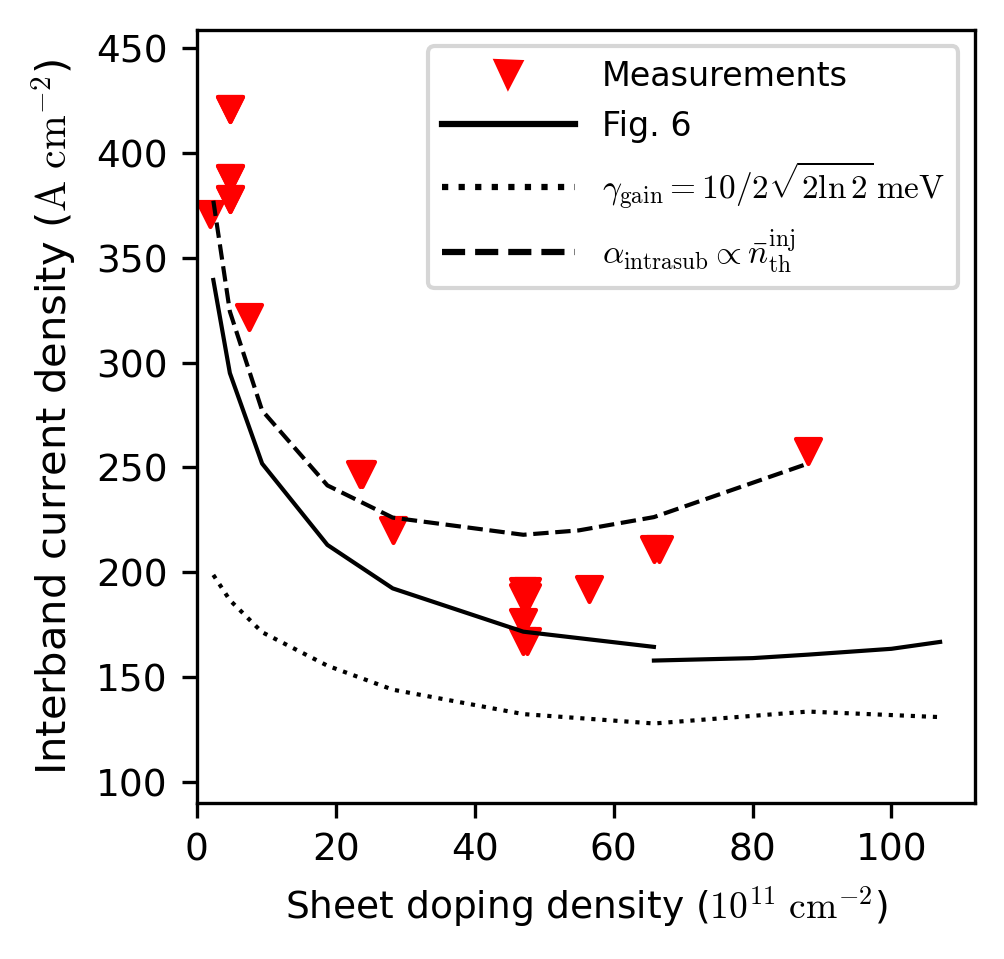}
	\else
		\includegraphics[width=\linewidth]{./images/paper/nextnanopy/GWJth_vs_sheetdoping_vs_intrasubband.png}
	\fi
    \caption{
		Auger recombination currents of the 4-well doping series simulated with a smaller $\gamma_\mathrm{gain}$ (dotted) and with the phenomenological model for the intrasubband absorption in the electron injector (dashed), compared with the result presented in Fig.\,\ref{fig:GWJth_vs_sheetdoping} (solid) and the measured threshold currents~\cite{Vurgaftman2011,WeihPhD,VurgaftmanPrivateComm} (triangles).
	}
	\label{fig:GWJth_vs_sheetdoping_vs_intrasubband}
\end{figure}

Table~\ref{table:simulation_parameters} summarizes the geometry and numerical parameters used in this study.
The mirror reflectivity $R$ was derived from the average refractive index $n_\mathrm{r} \approx 3.5$ of the cascade period.
The optical confinement factor $\Gamma$ was calculated by a Helmholtz equation solver.
The waveguide loss was obtained by the Drude model and a mobility model that depends on the carrier concentration~\cite{BorislavInAsGaSbComparison}.

Fig.\,\ref{fig:GWJth_vs_sheetdoping_vs_intrasubband} shows the influence of the parameters $\gamma_\mathrm{gain}$ and $\alpha_\mathrm{intrasub}$ on $J_\mathrm{A}$ in Fig.\,\ref{fig:GWJth_vs_sheetdoping}. The former depends on scattering strengths, while the latter emulates the intrasubband absorption in the electron injector in the threshold condition as
\begin{equation}\label{eq:gain_match_loss_intrasub}
	\Gamma
	\left(
		g_\mathrm{th} 
		-
		\alpha_\mathrm{intrasub}
	\right)
	=
	\frac{1}{L_\mathrm{cav}}\ln{\left(\frac{1}{R}\right)}
	+
	\alpha_\mathrm{w}.
\end{equation}
$\alpha_\mathrm{intrasub}$ was obtained as follows.
We first found at $N_\mathrm{D} = \qty{8.8e12}{\per\cm\squared}$ that, if $\alpha_\mathrm{intrasub}$ were about \qty{29}{\per\cm}, $J_\mathrm{A}$ would be \qty{252}{\ampere\per\cm\squared}. When the contribution from the spontaneous emission is \qty{6}{\ampere\per\cm\squared} and the SRH recombination is negligible, this sums up to the measured $J_\mathrm{th} = \qty{258}{\ampere\per\cm\squared}$. If $\alpha_\mathrm{intrasub}$ is proportional to the electron density in the electron injector, we have $\alpha_\mathrm{intrasub}~(\unit{\per\cm}) = \num{7.9e-18}~\bar{n}_\mathrm{th}^\mathrm{inj}~(\unit{\per\cm\cubed})$. Using $\bar{n}_\mathrm{th}^\mathrm{inj}$ in Fig.\,\ref{fig:comparison_doping_vs_In}(b2), we searched for the new thresholds by Eq.\,\eqref{eq:gain_match_loss_intrasub} and calibrated $J_\mathrm{A}$ of the 4-well doping series.

\section{Oscillator strengths}
\label{sec:oscillator_strengths}
While the envelope function overlap represents a major part of the oscillator strength of ICL lasing levels, the net material gain also involves parasitic absorptions inside the conduction or valence bands. Therefore, the envelope overlap alone is not sufficient to evaluate the net gain. 
Calculation of the oscillator strengths requires a matrix representation of the dipole operator along the $z$ axis and the momentum operator in the $x$--$y$ plane. The energy eigenstates of $\bm{H}_\mathrm{RRS}$ 
are used as the basis for a better compatibility with the NEGF formalism.
The $z$ component is readily obtained from the diagonal matrix in the RRS basis, $\mathrm{diag}(z_1, z_2, \cdots)$.

For the in-plane directions $j \in \{x, y\}$, we opt for the momentum picture of the light-carrier interaction since the position operator $\hat{x}_j$ in the directions of translational invariance is ambiguous. The oscillator strengths calculated from the momentum and dipole pictures are equal if (i) the momentum operator coincides with the commutator $(m_0/\imag\hbar)[\hat{x}_j, \hat{H}]$, and (ii) the oscillator strength is represented in the eigenbasis of $\hat{H}$~\cite{BetheSalpeter}. The requirement (i) is not straightforward when the Hamiltonian contains spin-orbit coupling, second-order \kp coupling via remote bands~\cite{Dupont2024} as in Eq.\,\eqref{eq:definition_S}, or position-dependent parameters.
Nevertheless, we retain consistency with the growth direction by \textit{defining} the in-plane momentum operator by $\hat{p}_j \equiv (m_0/\imag\hbar)[\hat{x}_j, \hat{H}]$. 
For homogeneous infinite lattices, the \kp perturbation theory~\cite{Luttinger1954,Szmulowicz1995} shows that
\begin{equation}\label{eq:finite_band_momentum_operator}
	\frac{m_0}{\imag\hbar}
	\left[
		\hat{x}_j, \hat{H}_\mathrm{kp}(\mathbf{k})
	\right]_{\nu\nu'}
	=
	\frac{m_0}{\hbar}
	\left[
		\frac{\partial \hat{H}_\mathrm{kp}(\mathbf{k})}{\partial k_j}
	\right]
	_{\nu\nu'}
	+
	\mathcal{O}(k^2),
\end{equation}
where $\nu$ and $\nu'$ enumerate the \kp bands considered in the model. We neglect the second and higher order terms according to the accuracy of the Hamiltonian \eqref{eq:8kp_Hamiltonian}.
For heterostructures, it is tempting to assign Eq.\,\eqref{eq:finite_band_momentum_operator} to each real-space grid point with position-dependent material parameters. However, the confinement along the growth direction (derivatives $\mathrm{d}/\mathrm{d}z$) affects $\hat{p}_{x,y}$ via the products $k_xk_z$ and $k_yk_z$ in the Hamiltonian \eqref{eq:H_vv}.
We instead insert the heterostructure Hamiltonian [Eq.\,\eqref{eq:8kp_Hamiltonian_hetero}] after discretization into Eq.\,\eqref{eq:finite_band_momentum_operator} to obtain the matrix representation in the tensor product basis of the \kp bands and real-space coordinates (enumerated by $s,s'$),
\begin{equation}\label{eq:heterostructure_momentum_operator}
	\left(
		\hat{p}_j
	\right)_{ss'}
	=
	\frac{m_0}{\hbar}
	\frac{\partial \bm{H}_{ss'}^{8N \times 8N}(\mathbf{k}_\parallel)}{\partial k_j}.
\end{equation}
The matrix is transformed first to the RRS basis and then to the eigenbasis of $\bm{H}_\mathrm{RRS}$, and is substituted to Fermi's golden rule.

Quantum confinement and selection rules at each $\mathbf{k}_\parallel$ point are encoded in these matrix elements. 
An optimally designed W-QW is asymmetrical and the oscillator strength between the lasing levels reaches the maximum approximately at the threshold electric field. However, the relative increase in an asymmetric W-QW design is much milder than e.g.\ for triple electron well designs~\cite{Dyksik2017,Petrovic2025_VQW}. We thus neglect the bias-dependence in Eq.\,\eqref{eq:heterostructure_momentum_operator}.

Consequently, the oscillator strength between two energy eigenstates at the in-plane momentum $\mathbf{k}_\parallel$ is given by 
\begin{align}
	O_{nm}(\polarizationVector, \mathbf{k}_\parallel)
	=&
	\frac{\hbar}{m_0^2\omega_{nm}}
	\sum_{j=x,y}
	\left|
		e_j 
		\langle n |
			\hat{p}_j(\mathbf{k}_\parallel)
		| m \rangle
	\right|^2 \notag\\
	&+
	\hbar\omega_{nm}
	\left|
		e_z
		\langle n |
			\hat{z}
		| m \rangle
	\right|^2,
\end{align}
where $\hbar\omega_{nm}$ is the energy separation of the two states.

\section{Carrier-carrier scattering}
\label{app:carrier_carrier_scattering}
The Auger process is an inelastic carrier-carrier scattering and therefore requires an approach beyond the Hartree--Fock approximation.
Hedin~\cite{Hedin1965} proposed a perturbation expansion of Green's functions around the screened Coulomb interaction $\bm{W}$. Since $\bm{W}$ describes the effective (dressed) interaction, approximations based on this expansion are expected to describe the electron-electron interaction in a physically transparent manner. 
The screened Coulomb interaction defined on the Keldysh contour obeys Dyson's equation~\cite{Stefanucci2025},
\begin{widetext}
	\begin{align}
		W_{\alpha\beta}(1;2) 
		&= 
		V_{\alpha\beta}(1;2) 
		+
		\sum_{\gamma\delta}
		\int\spaced{3}\int\spaced{4} 
		V_{\alpha\gamma}(1;3) \Pi_{\gamma\delta}(3;4) W_{\delta\beta}(4;2), \label{eq:W_Dyson}
	\end{align}
\end{widetext}
where $1 = (x_1,y_1,t_1)$ is the spacetime coordinates of an electron and the indices $z_i$ label the RRS modes. The bare Coulomb interaction is instantaneous, i.e., $\bm{V}(1;2) \propto \delta(t_1 - t_2)$. Note that $\bm{V}$, $\bm{W}$, and $\bm{\Pi}$ have two instead of four mode indices as the scattering from one RRS mode to another is negligible since the RRS basis diagonalizes the reduced position operator~\cite{Thygesen2008,Grange2014}.
The polarization function $\bm{\Pi}$ describes the excitation of the \quotes{partner} electron that interacts with the electron of interest. 

The screened Coulomb interaction affects the electron of interest via the self-energy $\bm{\Sigma}$. It consists of diagrams involving the Green's function and $\bm{W}$. The $GW$ approximation is to approximate $\bm{\Sigma}$ and $\bm{\Pi}$ in Eq.\,\eqref{eq:W_Dyson} by the lowest-order diagrams~\cite{Stefanucci2025}
\begin{align}
	\Sigma_{\alpha\beta}(1;2)
	&\approx
	\imag G_{\alpha\beta}(1;2) 
	W_{\alpha\beta}(1;2), \label{eq:GW_self-energy_diagram} \\
	\Pi_{\alpha\beta}(1;2)
	&\approx
	-G_{\alpha\beta}(1;2) 
	G_{\beta\alpha}(2;1), \label{eq:polarization_diagram}
\end{align}
which account for inelastic carrier-carrier scattering. 

The greater ($>$) component of Eq.\,\eqref{eq:W_Dyson} is calculated assuming translational symmetry in the in-plane directions and steady states,
\begin{align}
	\bm{W}_{\mathbf{q}_\parallel}^>(E)
	&=
	\bm{W}_{\mathbf{q}_\parallel}^\mathrm{R}(E)
	\bm{\Pi}_{\mathbf{q}_\parallel}^>(E)
	\bm{W}_{\mathbf{q}_\parallel}^\mathrm{A}(E).
	\label{eq:W_greater_general}
\end{align}
The retarded ($\mathrm{R}$) component of Eq.\,\eqref{eq:W_Dyson} is calculated considering a three-dimensional interacting electron gas in a medium with the static dielectric constant $\epsilon_0\epsilon(0)$. The matrices reduce to scalar functions in this case and
\begin{widetext}
	\begin{align}
		W_\mathbf{q}^\mathrm{R}(t_1; t_2) 
		&=
		V_\mathbf{q}(t_1; t_2)
		+
		\int\spaced{t_3} 
		\int\spaced{t_4}
		V_\mathbf{q}(t_1; t_3)
		\Pi_\mathbf{q}^\mathrm{R}(t_3; t_4)
		W_\mathbf{q}^\mathrm{R}(t_4; t_2), \label{eq:W_R_general}
	\end{align}
\end{widetext}
where $V_\mathbf{q}(t_1; t_2) = \delta(t_1 - t_2) e^2 / (\epsilon_0\epsilon(0)L^3 q^2)$ is the bare Coulomb potential in the reciprocal space with the normalization volume $L^3$.
We further make the quasi-static screening approximation, i.e., assume that the electron gas is not too dense so that carriers rearrange themselves in response to an addition of an electron much faster than getting scattered by the environment~\cite{BinderKoch_dynamics}. $W_\mathbf{q}^\mathrm{R}$ becomes instantaneous and Eq.\,\eqref{eq:W_R_general} can be solved algebraically,
\begin{equation}\label{eq:W_R_algebraic}
	W_\mathbf{q}^\mathrm{R}
	=
	\frac{V_\mathbf{q}}{1 - V_\mathbf{q}\Pi_\mathbf{q}^\mathrm{R}}.
\end{equation}
This can be seen as a renormalized Coulomb interaction with the effective screening length
\begin{equation}\label{eq:screening_length_3DEG}
	\lambda_\mathrm{s}(\mathbf{q})
	=
	\left(
		- 
		q^2 V_\mathbf{q}
		\Pi_\mathbf{q}^\mathrm{R}
	\right)^{-1/2}.
\end{equation}
Similarly to Eqs.\,\eqref{eq:polarization_lessgtr_in_energy_c} and \eqref{eq:polarization_lessgtr_in_energy_v}, we approximate $\Pi_\mathbf{q}^\mathrm{R} \approx \Pi_\mathbf{0}^\mathrm{R}$.
To apply the resulting interaction,
\begin{equation}
	W_\mathbf{q}^\mathrm{R} 
	= 
	\frac{V_\mathbf{q}}{1 + (q\lambda_\mathrm{s})^{-2}}
	=
	\frac{e^2}{\epsilon_0\epsilon(0)L^3}
	\frac{1}{q^2 + \lambda_\mathrm{s}^{-2}},
\end{equation}
to QWs, we Fourier-transform $q_z$ to $z$ and obtain
\begin{subequations}\label{eq:W_RA_quasistatic_electron_gas}
	\begin{align}
		W_{\mathbf{q}_\parallel}^\mathrm{R}(z)
		&=
		\frac{e^2}{2\epsilon_0\epsilon(0)L^2}
		\frac{\ee{-q_\mathrm{s} |z|}}{q_\mathrm{s}}, \\
		W_{\mathbf{q}_\parallel}^\mathrm{A}
		&=
		\left( W_{\mathbf{q}_\parallel}^\mathrm{R} \right)^\dagger
		=
		W_{\mathbf{q}_\parallel}^\mathrm{R},
	\end{align}
\end{subequations}
where $q_\mathrm{s} = \sqrt{q_\parallel^2 + \lambda_\mathrm{s}^{-2}}$.

The evaluation of the screening length in Eq.\,\eqref{eq:screening_length_3DEG} is involved in general because the retarded polarization has a complex dependence on the subband occupation. 
As a first estimate, we deduce the retarded polarization by comparing the denominator of Eq.\,\eqref{eq:W_R_algebraic} with the long-wavelength, static, and quasi-equilibrium limit of the Lindhard dielectric function~\cite{HaugKoch,BinderKoch_dynamics}. The carrier temperature is set to \qty{300}{\kelvin} as in the gain calculation. 
The resulting screening length ranges from \qtyrange{1.8}{3.0}{\nm} for the designs in Table~\ref{table:design_variations} and is of the same order of magnitude as the value used in this work, $\lambda_\mathrm{s} = \qty{4.0}{\nm}$.

The Lindhard function in the static limit includes the Debye--H\"{u}ckel and Thomas--Fermi screening models as the non-degenerate (Maxwell--Boltzmann) and degenerate limiting cases, respectively. While the Debye screening length can be used in type-II superlattice photodetectors, ICL simulations may be more accurate with the Thomas--Fermi screening~\cite{Kolek2024_NUSOD}.

\ifFlatFolderStructure
	\bibliography{Refs}
\else
	\bibliography{../references/Refs}
\fi

\end{document}